\documentclass{aa}

\usepackage{graphicx}
\usepackage{txfonts}

\usepackage{natbib}

\usepackage{soul}
\usepackage{color}
\usepackage[colorlinks=true,citecolor=blue]{hyperref}
\begin{document}

\title{High-resolution transmission spectroscopy study of ultra-hot Jupiters HAT-P-57b, KELT-17b, KELT-21b, KELT-7b, MASCARA-1b, and WASP-189b}

\titlerunning{Atmospheres of 6 UHJs}  

\author{M. Stangret\inst{1}\fnmsep\inst{2}, N. Casasayas-Barris\inst{3}, E. Pallé\inst{1}\fnmsep\inst{2},  J. Orell-Miquel\inst{1}\fnmsep\inst{2}, G. Morello\inst{1}\fnmsep\inst{2}, R. Luque\inst{4},  G. Nowak\inst{1}\fnmsep\inst{2}, F. Yan\inst{5} }
   
\authorrunning{M. Stangret et al.}

   \institute{Instituto de Astrofísica de Canarias, Vía Láctea s/n, 38205 La Laguna, Tenerife, Spain \\
             \email{mstangret@iac.es}          
             \and 
             Departamento de Astrofísica, Universidad de La Laguna, 38200 San Cristobal de La Laguna, Spain
             \and 
             Leiden Observatory, Leiden University, Postbus 9513, 2300 RA Leiden, The Netherlands
             \and
             Instituto de Astrofísica de Andalucía (IAA-CSIC), Glorieta de la Astronomía s/n, 18008 Granada, Spain
             \and
             Institut f\"ur Astrophysik, Georg-August-Universit\"at, Friedrich-Hund-Platz 1, 37077 G\"ottingen, Germany
             }

   \date{Received 2021; accepted  2021}

 
  \abstract
 {

Ultra-hot jupiters (UHJs) are giant planets on short orbital periods with high equilibrium temperature ($T_{\mathrm{eq}}$) values. Their hot, extended atmospheres are perfect laboratories for transmission spectroscopy studies based on high-resolution spectrographs. 
In recent years, a variety of atoms and molecules were found in their atmospheres, using different methods such as cross-correlation or transmission and emission spectroscopy. Here, we present the studies of six ultra-hot Jupiters: HAT-P-57b, KELT-7b, KELT-17b, KELT-21b, MASCARA-1b, and WASP-189b, based on high-resolution observations obtained with HARPS-N and HARPS spectrographs. By applying line and cross-correlation transmission spectroscopy methods, we searched for the absorption features of a broad range of atomic and molecular species. We did not detect any absorption features in our sample of UHJs, with the exception of WASP-189b, for which we detected \ion{Fe}{i}, \ion{Fe}{ii}, and \ion{Ti}{i} using cross-correlation. The transmission spectroscopy of single lines for WASP-189b revealed several absorption features (including H$\alpha$, H$\beta,$ and Ca H\&K), but they remain tentative pending a better modeling of the gravity darkening deformation of the Rossiter-McLaughlin effect.
The non-detections with regard to the rest of the planets can be explained via a combination of stellar pulsations and the Rossiter-McLaughlin effect, which mask possible planetary signals for most of the planets, and by the low signal-to-noise ratios (S/N) of the observations for KELT-21b. Here, we compare our results with the known population of planets for which atmospheric detections have been reported in the literature. We find that the empirical frontier between hot and ultra-hot planets, based on the detection of atomic and ionized species in their atmospheres, can be established as $T_{eq} = 2150 K$.

}


   \maketitle
 
%
\section{Introduction}

High-resolution spectroscopic observations are capable of detecting and characterizing exoplanetary atmospheres by exploiting the differential radial velocities of the planet, the host star, and the Earth. The most suitable planets for the application of this technique with the current instrumentation are hot Jupiters (HJs), which are planets with high-temperatures, short orbital periods and extended atmospheres. In addition, ultra-hot Jupiters (UHJs), namely, giant planets with equilibrium temperatures $> 2000 K$ ($T_{\mathrm{eq}}$) \citep{Parmentier2018}, are of particular interest. Their short orbital periods and the strong irradiation received from their host star, along with their tidally locked systems, produce significant differences between the day- and night-side chemical compositions of their atmospheres.

Most UHJs are ideal targets for both transmission and emission spectroscopy thanks to their extended atmospheric scale-heights and brightness.
Theoretical simulations indicate that clouds are unlikely to form in the day-side atmospheres of UHJs, as the high temperatures prevent condensation \citep{helling_wasp_18,helling_2021_cpt} -- with the exception of highly refractory species such as Al and Ti \citep{wakeford_2017}.
Hence, the observed atmospheric chemistry should be representative of the bulk composition, which can be used to constrain their possible formation scenarios and evolutionary pathways \citep{turrini_2021}.
Depending on the vertical temperature profile, most species may appear in their ionised state and molecules can be partially or fully dissociated \citep{Parmentier2018, mansfield_2020}.
Previous studies have also suggested that the upper atmospheres of UHJs may present a thermal inversion, for example, in the atmosphere of WASP-33b \citep{2015Haynes_wasp33, nugroho_2020_wasp33_FeI,  2021_Cont_wasp-33}, WASP-121b \citep{evans_2016,evans_2018}, WASP-103b \citep{2018_Kreidberg_wasp103}, WASP-18b \citep{2017_Sheppard_wasp18}, HAT-P-7b \citep{2016_Armstrong_hatp7}, KELT-9b \citep{Pino_2020_Kelt:9}, and WASP-189b \citep{Yan_2020_wasp-189}.

With regard to atmospheric species, a variety of atoms and ions have been detected in UHJ atmospheres. For KELT-9b \citep{gaudi_kelt_9b}, which is by far the hottest UHJ known to date, significant detections have been reported including: a H$\mathrm{\alpha}$ line, revealing an extended hydrogen atmosphere \citep{YanKELT9}; \ion{Cr}{ii}, \ion{Fe}{i}, \ion{Fe}{ii}, \ion{Mg}{ii}, \ion{Na}{i}, \ion{Sc}{ii}, \ion{Ti}{ii,} and \ion{Y}{ii}; as well as evidence of \ion{Ca}{i}, \ion{Cr}{i}, \ion{Co}{i,} and \ion{Sr}{ii} in its atmosphere \citep{Hoeijmakers_2018_kelt9, Hoeijmakers_2019_kelt9}, along with a H and \ion{Mg}{i} triplet \citep{Cauley_2019_kelt9}. Many of these species have also been detected in other much colder UHJs: MASCARA-2b/KELT-20b \citep{Casasayas2018, Casasayas2019, Stangret_2020_MASCARA-2, Nugroho2020_KELT20, Hoeijmakers_mascara2}, WASP-12b \citep{Jensen_2018_wasp12}, WASP-33b \citep{Nugroho_2017_wasp33, yan-2019-calcium-kelt9-wasp33, nugroho_2020_wasp33_FeI, Yan_2021_wasp33}, WASP-76b \citep{seidel-2019-wasp-76, Ehrenreich_wasp76}, WASP-121b \citep{2020Cabot_wasp-121, 2020BenYami_wasp121,2020arXiv200106430G, 2020HoeijmakersHEARTS,2021_Borsa_wasp121}, and WASP-189b \citep{Yan_2020_wasp-189}.

Additionally, while molecules are expected to be heavily dissociated, H$_2$O has been detected in the atmospheres of WASP-19b \citep{huitson_2013}, WASP-12b \citep{kreidberg_2015_wasp12b}, WASP-121b \citep{evans_2016,mikal-evans_2020}, WASP-76b \citep{tsiaras_2018,edwards_2020}, and, tentatively, on KELT-7b \citep{pluriel_2020}. There have also been some debatable detections of either TiO and VO in WASP-33b \citep{Nugroho_2017_wasp33,serindag_2021}, WASP-19b \citep{sedaghati_2017,sedaghati_2021}, and WASP-76b \citep{tsiaras_2018,edwards_2020}.
If confirmed, TiO and VO could partially explain the observed thermal inversions in UHJ atmospheres \citep{hubeny_2003,fortney_2008}.

In this paper, we present studies of the atmospheric composition of six UHJs: HAT-P-57b, KELT-7b, KELT-17b, KELT-21b, MASCARA-1b, and WASP-189b.  We used high-resolution spectroscopic observations, as described in Section \ref{sec:Observations}, using single lines transmission spectroscopy method, discussed in Section \ref{sec:TS}, and the cross-correlation method, discussed in Section \ref{sec:CC}.
In Section \ref{sec:Results} we report the results of our analysis. We note that such phenomena as pulsation, center-to-limb variation (CLV), and Rossiter-McLaughlin (RM) effects prevented us from fully exploring the atmospheric composition of these six UHJs. In Section \ref{sec:Discussion}, we discuss our findings and set them in context with other HJs and UHJs whose atmospheres have been studied before.


\section{Observations}\label{sec:Observations} 
We analyzed observations taken using the HARPS-North spectrograph \citep{Cosentino_2012_harpsn}, mounted at the 3.6 m Telescopio Nazionale Galileo (TNG) at Observatorio del Roque de los Muchachos (ORM) in La Palma, Spain, as well as observations carried out with the HARPS spectrograph \citep{Mayor_2003_HARPS}, mounted at the ESO La Silla 3.6 m telescope in Chile. We analyzed data from six UHJs: HAT-P-57 b, KELT-7b, KELT-17 b, KELT-21 b, MASCARA-1 b, and WASP-189 b. All stellar and planetary parameters for these six planets are presented in Table \ref{table:param} and an observation log summary is presented in Table \ref{table:log}

\begin{table*}[h]
\caption{Stellar and planetary  parameters for HAT-P-57 adopted from \citet{Hartman_hat-p-57}, KELT-7 adopted from \citet{Beryla_Kelt-7} (a) from Gaia Archive, KELT-17 adopted from \citet{Zhou_kelt-17}, KELT-21 adopted from  \citet{Johnson_kelt-21}, MASCARA-1 adopted from \citet{talens_mascara-1}, and WASP-189 adopted from \citet{2020_Lendl_wasp189} and (b) from \citet{Anderson_wasp-189}.  }             
\label{table:param}      
\centering          
\resizebox{\textwidth}{!}{\begin{tabular}{l c c c c c c c c c  }     
\hline\hline       
Description&HAT-P-57  & KELT-7 & KELT-17 & KELT-21& MASCARA-1& WASP-189\\ 
\hline     
    & & &  &  &  &  &  &  \\ [-0.7em] 

 V-band magnitude & 10.465 $\pm$ 0.029 & 8.540 $\pm$  0.030 & 9.286 $\pm$ 0.051 & 10.48 $\pm$ 0.04 & 8.3&  6.64\tablefoottext{b}\\  
 
Projected rotation speed [km/s]& 102.1 $\pm$ 1.3 &65.0$^{+6.0}_{-5.9}$& 44.2$^{+1.5}_{-1.3}$ & 146.03 $\pm$ 0.48   &109.0 $\pm$ 4  & 93.1 $\pm$ 1.7\\[0.15em]

Surface gravity log [cgs]& 4.251 $\pm$ 0.018 & 4.149 $\pm$ 0.019 &4.220$^{+0.022}_{-0.024}$ & 4.173$^{+0.015}_{-0.014}$  & 4 &3.9 $\pm$ 0.2 \\[0.15em]

Metallicity & -0.25 $\pm$ 0.25 & 0.139$^{+0.075}_{-0.081}$ &-0.018$^{+0.074}_{-0.072}$  & -0.405$^{+0.032}_{-0.033}$ &  0&+0.29$\pm$ 0.13 \\[0.15em]

Stellar mass [$M_{\sun}$] & 1.47 $\pm$ 0.12 & 1.535$^{+0.066}_{-0.054}$ &1.635$^{+0.066}_{-0.067}$ & 1.458$^{+0.029}_{-0.028}$  & 1.72 $\pm$ 0.07 &2.030 $\pm$ 0.066 \\[0.15em]

Stellar radius  [$R_{\sun}$] & 1.50 $\pm$ 0.05 & 1.732$^{+0.043}_{-0.045}$& 1.645 $^{+0.060}_{-0.055}$&1.638 $\pm$ 0.034  & 2.1 $\pm$ 0.2  & 2.36 $\pm$ 0.030\\[0.15em]
\hline     

Planet mass  [$M_{\rm J}$] & <1.85 & 1.28 $\pm$ 0.18 &1.31$^{+0.28}_{-0.29}$ & <3.91 & 3.7 $\pm$ 0.9 &1.99$^{+0.16}_{-0.14}$ \\[0.1em]

Planet radius [$R_{\rm J}$]& 1.413 $\pm$ 0.054 &1.533$^{+0.046}_{-0.047}$& 1.525$^{+0.065}_{-0.060}$ &1.586$^{+0.039}_{-0.040}$ &   1.5 $\pm$ 0.3 & 1.619  $\pm$  0.021\\[0.15em]

Equilibrium temperature [K]&  2200 $\pm$ 76& 2048 $\pm$ 27 &2087$^{+32}_{-33}$ & 2051$^{+29}_{-30}$ & 2570$^{+50}_{-30}$ & 2690$\pm$\tablefoottext{b} 260\\[0.2em]
\hline

Right ascension &18 18 58.32  & 05 13 10.93 & 08:22:28.21& 20 19 12.004 & 21h10m12.37s & 15h 02m 44s. 86\\[0.15em]
Declination & +10 35 50. 3 & +33 19 05.40 & +13:44:07.2& +32 34 51 77 & +10 44 19.9 & - 03  01  5 2.9\\[0.15em]

Period [days] & 2.4652950 $\pm$ 0.0000032 &2.7347749 $\pm$ 0.0000039 &3.0801718 $\pm$ 0.0000053 &3.6127628 $\pm$  0.0000033  &2.148780 $\pm$ 0.000008   &2.7240330  $\pm$ 0.0000042\tablefoottext{b} \\

 Transit duration [days] &0.14578 $\pm$ 0.00080&0.14630$^{+0.00097}_{-0.00092}$ &0.1448$^{+0.0014}_{-0.0013}$  &0.17106$^{+0.00091}_{-0.00092}$  &4.05$\pm$ 0.03    &0.1813 $\pm$ 0.0011\tablefoottext{b} \\
 
 Ingress/egress duration [days] & 0.01325 $\pm$ 0.00054 & 0.01835$^{+0.00092}_{-0.00089}$ &0.0181$^{+0.0012}_{-0.0011}$ &0.01864 $\pm$ 0.00076  & - & 0.01289$\pm$ 0.0043\tablefoottext{b} \\
 
 Semi-major axis [AU]  & 0.0406 $\pm$ 0.0011 &0.04415$^{+0.00062}_{-0.00052}$& 0.04881$^{+0.00065}_{-0.00061}$ & 0.05224$^{+0.00035}_{-0.00034}$&   0.043 $\pm$ 0.005 & 0.05053 $\pm$ 0.00098\\[0.15em]
 
  Inclination [deg]  & 88.26 $\pm$ 0.85 &83.76$^{+0.38}_{-0.37}$& 84.87$^{+0.45}_{-0.43}$ &86.46$^{+0.38}_{-0.34}$   & 87$^{+2}_{-3}$ &84.03 $\pm$ 0.14 \\[0.15em]
  
   Systemic velocity [km/s] &-9.62  & 40.748405 $\pm$ 0.7995454\tablefoottext{a} &28.0$\pm$0.1 & -13 $\pm$ 1 &11.20 $\pm$  0.08 &-24.465 $\pm$ 0.012 \\[0.15em]
   
   Projected obliquity [deg] & -16.7 < l < 3.3 or 27.6 < l < 57.4  & 9.7$\pm$5.2 & -115.9 $\pm$ 4.1 & -5.6 $^{+1.7}_{-1.9}$ & 69.5 $\pm$ 4 & 86.4$^{+2.9}_{-4.4}$ \\[0.15em]
   
   $K_p$, Planetary RV semi-amplitude [km/s]  & 179.08 & 174.59 &171.70 & 156.57 &  217.41& 200.71\\
\hline      
\end{tabular}}
\end{table*}

\begin{table*}[h]
\caption{Summary of the transit observations.}             
\label{table:log}      
\centering          
\resizebox{\textwidth}{!}{\begin{tabular}{c c c c c c c c}     
\hline\hline       
Object & Night  & Date of observation & Start UT& End UT& Texp (s)& Nobs  & S/N@590\,nm$^{\rm(a)}$\\
[0.2em]
\hline 

HAT-P-57 b & 1 & 2019-06-23 & 21:39 & 05:09 & 600 & 44 & 49\\  
HAT-P-57 b  & 2   & 2019-06-28 & 21:09& 03:47 & 600 &39 & 54\\
KELT-7 b  & 1 & 2017-12-06 & 21:39& 04:20 & 900& 27 & 95\\
KELT-7 b  & 2  & 2017-12-17 & 20:17& 01:59& 866 &24 & 42\\
KELT-17 b  & 1  & 2019-01-23 & 21:16&  05:12& 600 &46 & 45\\
KELT-17 b  & 2 & 2019-01-26 & 21:26& 04:55 & 600 &39 & 56\\
KELT-21 b  & 1  & 2019-08-12 & 20:37& 03:54 & 800 &32 & 32\\
MASCARA-1 b  & 1 & 2017-07-04 & 22:52& 05:12 & 450& 49 & 105\\
MASCARA-1 b  & 2 & 2017-08-01 & 21:34& 04:18 & 450& 52 & 119\\
MASCARA-1 b  & 3 & 2019-08-08 & 21:10& 05:08 & 400& 68 & 105\\
MASCARA-1 b  & 4 & 2019-09-05 & 20:08& 03:16 & 400 &61 & 65\\
WASP-189 b$^{\rm(b)}$  & 1 & 2019-04-15 &2:20& 07:01 &  200&74 &169\\
WASP-189 b  & 2 & 2019-05-06 & 21:41& 04:45 & 200 &111 &158\\
WASP-189 b$^{\rm(b)}$  & 3 & 2019-05-15 & 00:31 & 07:15 & 200 &105 &122\\[0.2em]
\hline 
\end{tabular}}
\tablefoot{\tablefoottext{a}{Mean S/N for each night, calculated around 590\,nm (order of 53 for HARPS-N and 104 for HARPS).} \tablefoottext{b}{Data sets observed with HARPS, all other observations were performed with HARPS-N.}}
\end{table*}

First, HAT-P-57b \citep{Hartman_hat-p-57} is an UHJ with an equilibrium temperature of 2200 K, a radius of 1.413 R$_J$, and an upper mass limit of 1.85 M$_{J}$. The planet orbits a main sequence A8V star, which may be a variable star ($\gamma$  Dor-type variable, classical $\delta$-Scuti, or $\gamma$-Dor instability strips) with a period of 2.46 days. 
We observed two full transits of HAT-P-57 b over the nights of 23 June 2019 and 28 June 2019. During night 1, we took 44 exposures of 600 s with 23 out-of-transit and 21 in-transit spectra (with $\phi$ varying from -0.067 to +0.06, where $\phi$ is planet orbital phase), with an average signal-to-noise ratio (S/N) of 49 calculated around 590 nm. During night 2, we took 39 exposures of 600 s with 18 out-of-transit and 21 in-transit spectra ($\phi$=-0.047 to +0.065), with an average S/N of 54.

\begin{figure*}[h]
  \includegraphics[width=0.9\textwidth]{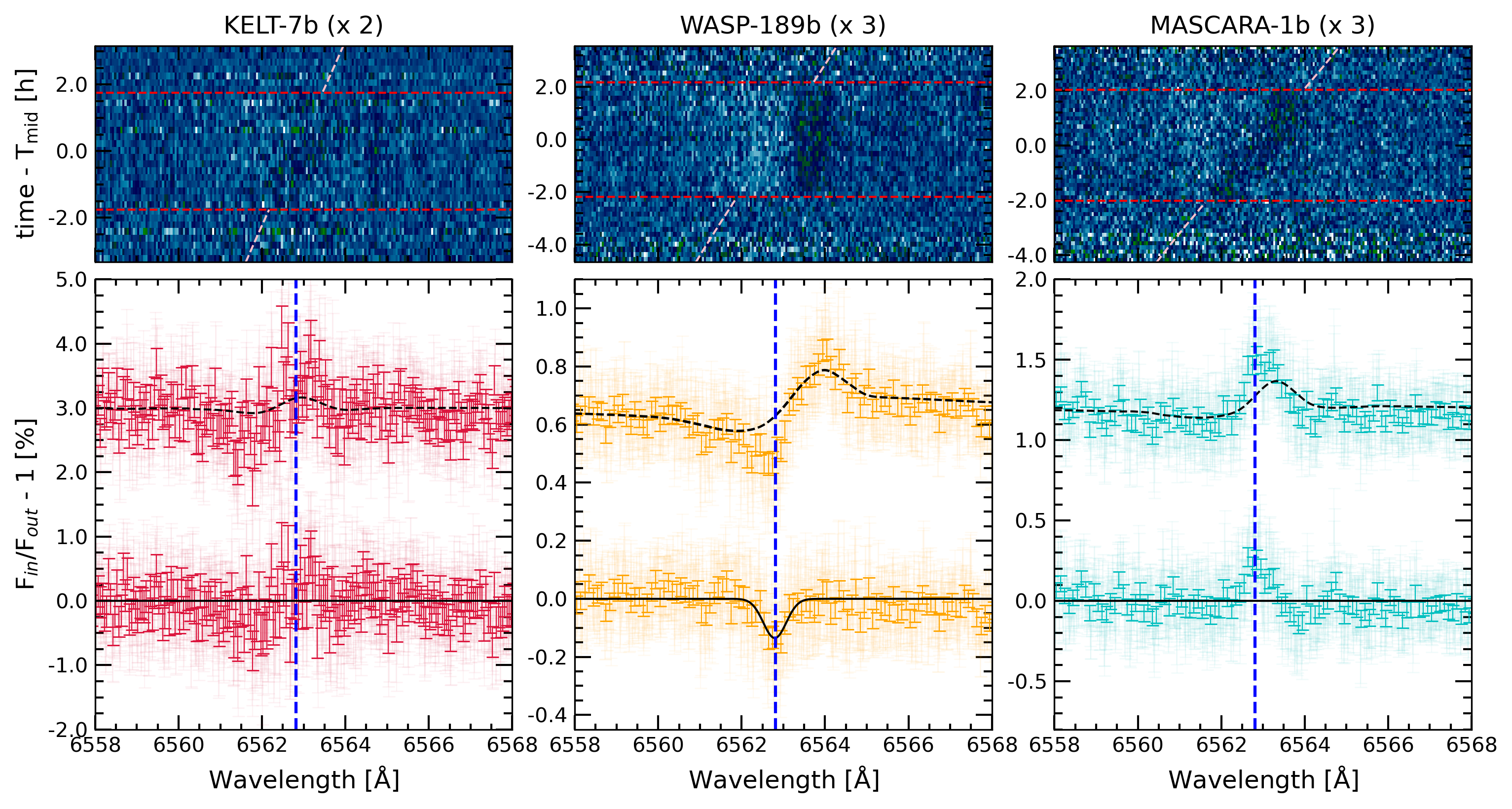}
  \includegraphics[width=0.9\textwidth]{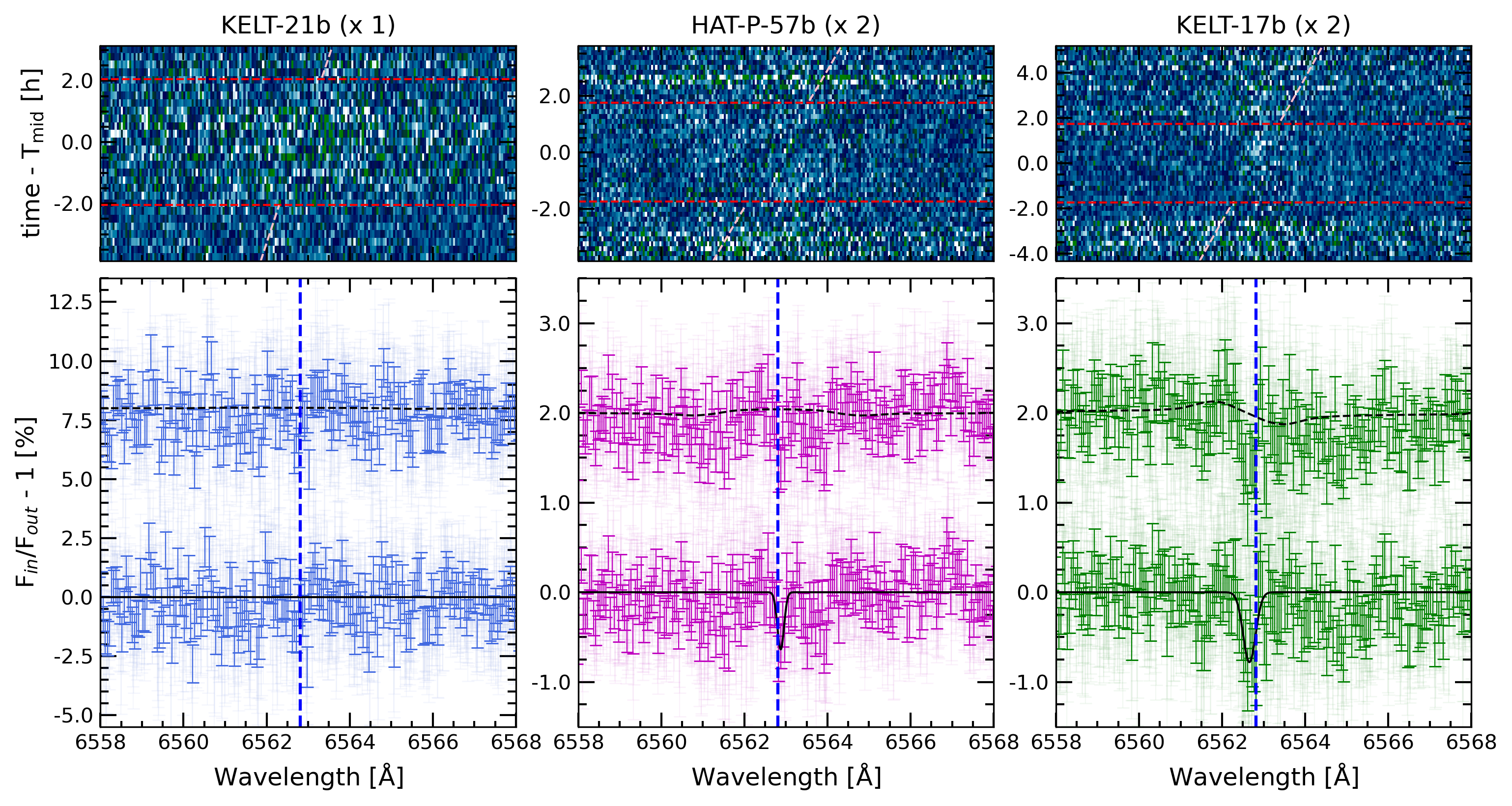}
  \caption{Transmission spectra around the H$\alpha$ line for the different UHJs analysed in this work (shown in different colors). First row: 2D map around H$\alpha$ of all nights combined showing the time evolution of the residuals after the spectra have been divided by the master out-of-transit spectrum. For better visualisation of the residuals, the maps include a wavelength binning of $0.05~{\rm \AA}$ and a 0.003 binning in orbital phase. Second row: In each panel, the top spectrum shows the transmission spectrum without correcting for the RM effect and the bottom spectrum shows the residuals after the correction. We note that the top spectrum has an offset on y-axis for a better visualization. In dashed lines we show the predicted RM effect, computed with the literature stellar and system parameters. In light color, we show the original data, and in solid color the data binned by $0.05~{\rm{\AA}}$. The dashed black line shows the RM model and the black line corresponds to the best fit Gaussian profile to the residuals near the H$\alpha$ position. The H$\alpha$ natural position is marked in a blue vertical-dashed line. KELT-21b, KELT-7b, and MASCARA-1b do not show any Gaussian-like residual to be fitted around H$\alpha$. The number close to the planet name in the title indicates the number of nights that are combined. We note the different y-axis scales of each panel due to the different S/N of the data. }
  \label{fig:Ha-all}
\end{figure*} 

KELT-7b \citep{Beryla_Kelt-7} is an UHJ with an equilibrium temperature of 2048 K, a radius of 1.533  R$_J$ and a mass of 1.28 M$_J$, orbiting an F-type star with a period 2.73 days. KELT-7b is a rapid rotating star ($v \sin{i}=76$~km\,s$^{-1}$). We observed one transit of KELT-7b during the night of 17 December 2017, where we took 24 exposures of 866 s, resulting in 9 out-of-transit and 15 in-transit spectra ($\phi$=-0.047 to +0.039), and with an average S/N of 42. In our analysis, we used one additional transit observations taken from archival data. The transit was observed during the night of 6 December 2017 (GAPS, Sozzetti), resulting in 27 exposures of 900~s, 12 out-of-transit, and 15 in-transit spectra ($\phi$=-0.049 to 0.053), with an average S/N of 95.

KELT-17b \citep{Zhou_kelt-17} is an UHJ with an equilibrium temperature of 2087 K, a radius of 1.525 R$_J$ and a mass of 1.31 M$_J$,  orbiting a main-sequence A-star with a period of 3.08 day. KELT-17 is a fast rotating star ($v \sin{i}=44.2$~km\,s$^{-1}$). We observed two full transits of KELT-17b during the nights of 23 January 2019 and 26 January 2019. During the first night, we took exposures of 600 s, resulting in 25 out-of-transit and 21 in-transit spectra ($\phi$=-0.032 to +0.075) and with an average S/N of 45. During the second night, we took 39 exposures of 600 s, resulting in 18 out-of-transit and 21 in-transit spectra ($\phi$=-0.055 to +0.045), with an average S/N of 56.

KELT-21b \citep{Johnson_kelt-21} is an UHJ with an equilibrium temperature of 2051 K, a radius of 1.586 R$_J$ and an upper limit mass of 3.91 M$_J$,  orbiting an A8V star with a period of 3.61 days. KELT-21 is a fast-rotating star with $v \sin{i}=146$~km\,s$^{-1}$. We observed one transit of KELT-21b during the night of 12 July 2019, when we took 32 exposures of 800 s, resulting in 13 out-of-transit and 19 in-transit spectra ($\phi$=-0.044 to +0.04), with an average S/N of 32.

MASCARA-1b \citep{talens_mascara-1} is an UHJ with an equilibrium temperature of 2570 K, a radius of 1.5 R$_J$ and a mass of 3.7 M$_J$, orbiting a fast-rotating A8 star ($v \sin{i}>100$~km\,s$^{-1}$) with a period of 2.15 days. We observed two full transits of MASCARA-1b during the nights of 8 August 2019 and 5 September 2019. On the first night, we took exposures of 450 s, resulting in 20 out-of-transit and 32 in-transit spectra ($\phi$=-0.059 to 0.071), with an average S/N of 119. On the second night we took 68 exposures of 400 s, resulting in 33 out-of-transit and 25 in-transit spectra ($\phi$=-0.082 to 0.072), with an average S/N of 105. Additionally, we used two transit observations from archival data. The first archival transit was observed during the night of 4 July 2017 (A35TAC\_26, Albrecht), resulting in 49 exposures of 450 s, 18 out-of-transit, and 31 in-transit ($\phi$=-0.065 to +0.0577), with an average S/N of 105. The second transit was observed during the night of 1 August 2017(A35TAC\_26,Albrecht), resulting in 61 exposures of 400 s, 25 out-of-transit, and 36 in-transit spectra ($\phi$=-0.071 to +0.067), with an average S/N of 65.

WASP-189b \citep{Anderson_wasp-189} is an UHJ with an equilibrium temperature of 2641 K, a radius of 1.374 R$_J$, and a mass of 2.13 M$_J$, orbiting a rapidly rotating A6 star ($v \sin{i}=100$~km\,s$^{-1}$) with a period of 2.72 days on a polar orbit ($\lambda$=89.3 $\deg$).

We observed one transit of WASP-189b during the night of 12 July 2019, when we took 111 exposures of 200 s, resulting in 43 out-of-transit and 68 in-transit spectra ($\phi$=-0.053 to +0.055), with an average S/N of 125.2. Additionally, we used two transit observations from archival data from HARPS. The first archival transit was observed during the night of 15 April 2019 (0103.C-0472, Hoeijmakers), resulting in 74 exposures of 200 s, 25 out-of-transit, and 49 in-transit ($\phi$=-0.059 to +0.013), with an average S/N of 169. The second transit was observed during the night of 15 May 2019 (0103.C-0472, Hoeijmakers), resulting in 105 exposures of 200 s, 40 out-of-transit, and 65 in-transit ($\phi$=-0.073 to +0.03), with an average S/N of 122.

Observations were reduced with the HARPS-South and HARPS-North data reduction software (DRS, \citet{DRS1, DRS2}). We used version 3.7, which extracts the spectra order-by-order and performs flat-field using daily calibration set and combines all the orders to create a one-dimensional (1D) spectrum.

\section{Methods}\label{sec:methods}

\subsection{Transmission spectroscopy of single lines}\label{sec:TS} 

We searched for atmospheric absorption features from each of the six ultra-hot Jupiter planets. We followed the now common methodology presented in \citet{Wytt2015} and \citet{Casasayas2018}, however, we first corrected the spectra for telluric contamination using {\tt Molecfit} \citep{Molecfit1,Molecfit2} following \citet{Allart2017}. Here, the master out-of-transit spectrum is calculated as the weighted mean of all out-of-transit spectra, using the S/N of each spectrum as weight. The transmission spectrum is calculated as the combination of the results between the first and last contact of the transit in order to achieve a slightly higher S/N.

All six planets studied here orbit fast-rotating stars ($v\sin i > 44~{\rm km s^{-1}}$) and as a result, the RM effect is expected to be strong. The RM effect, as well as the CLV, strongly affect transmission spectroscopic observations and it is important to remove these effects before computing the final result. 
In order to correct for these effects, we followed the same methodology described in \citet{Casasayas2019}. First, we modeled the stellar spectra in two different ways: one containing the RM and CLV effects and one containing RM effect only, at different planet orbital phases. These stellar models were computed with the \texttt{Spectroscopy Made Easy} tool (SME, \citet{SME}), using the Kurucz ATLAS9 \citep{ATLAS92003} and VALD3 line list \citep{VALD3}. The stellar spectra were modeled for different limb-darkening angles assuming local thermodynamical equilibrium (LTE) and solar abundance. In all cases, we adopted the system parameters that are presented in Table \ref{table:param}. While calculating stellar models, we took into consideration the fact that the planet covers different stellar regions at each orbital phase. In the final step, we divided the synthetic stellar models at each phase by the stellar models computed when the planet is not transiting.

The planets in our sample orbit very hot stars ($T_{\rm eff}\gtrsim7000~{\rm K}$) and, thus, the center-to-limb variation is expected to be fainter than in cooler stars \citep{CzeslaCLV2015}. Given that the S/N achieved in single-line transmission spectroscopy is not high enough to distinguish this contribution for a single line, we only considered the RM contribution here. The models containing the CLV effect are relevant to the part of the study detailed in Section \ref{sec:CC}.

To uncover possible planetary atmospheric signals, we subtracted the RM model from the final transmission spectrum. The models have their own uncertainties, which emerge from system parameters and stellar atmospheric models uncertainties. Thus, when subtracting the RM effect, we could be over- or undercorrecting the overall spectral shape. However, the estimated RM effect for these planets is relatively broad in wavelength, with amplitudes of more than $1~{\rm \AA}$ (with positive or negative relative flux changes depending on the system geometry; see Figure~\ref{fig:Ha-all}). If the expected exoplanet absorption has a contrast that could be observable with the S/N achieved in the observations and has a full-width at half maximum (FWHM) narrower than this RM shape, we would be able to observe a deviation between the model and the data.  The results following the subtraction of the RM model from the data are presented in Figure~\ref{fig:Ha-all}.

\begin{figure*}[h]
  \includegraphics[width=0.9\textwidth]{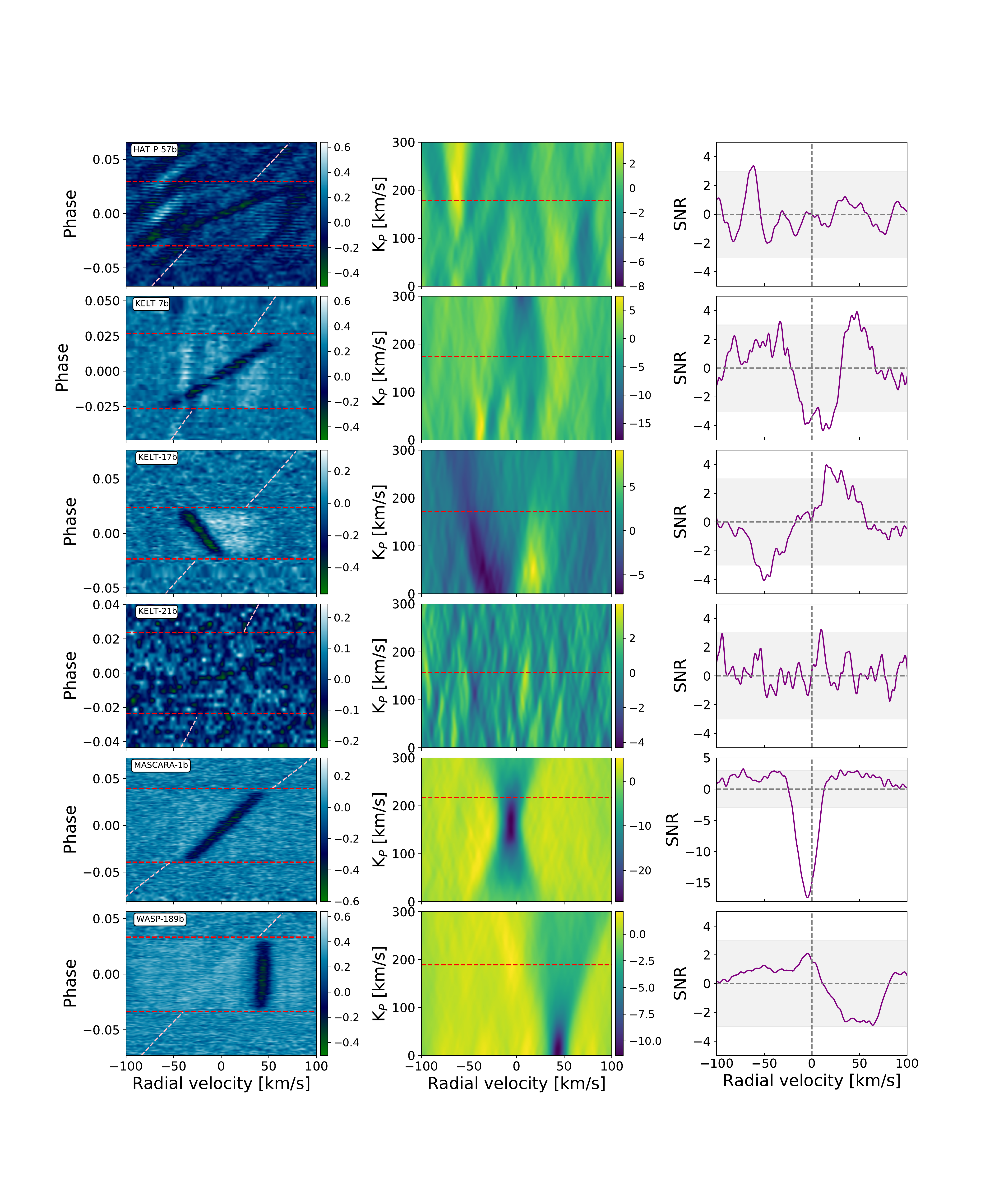}
  \caption{Cross-correlation results of \ion{Fe}{i} lines for the six UHJs. For each of the studied planets, we present the following plots. Left:\ Residuals map for \ion{Fe}{i}, where the white tilted line represents the expected velocity of the planet and red horizontal lines show the beginning and the end of the transit, region where we expect the signal from the planet; center map: $K_p$ map for $K_p$ in the range of 0 to 300~km\,s$^{-1}$ and the red line represents the theoretical $K_p$ value for the planet. We expect the signal to be coming from a planet at radial velocity of 0~km\,s$^{-1}$ and theoretical $K_p$. Right: S/N plot for expected $K_p$ value, the grey vertical lines represent 0~km\,s$^{-1}$ and gray region represents S/N between -3 and 3. }
  \label{fig:FeI-all_RM}
\end{figure*}

\begin{figure*}[h]
  \includegraphics[width=0.9\textwidth]{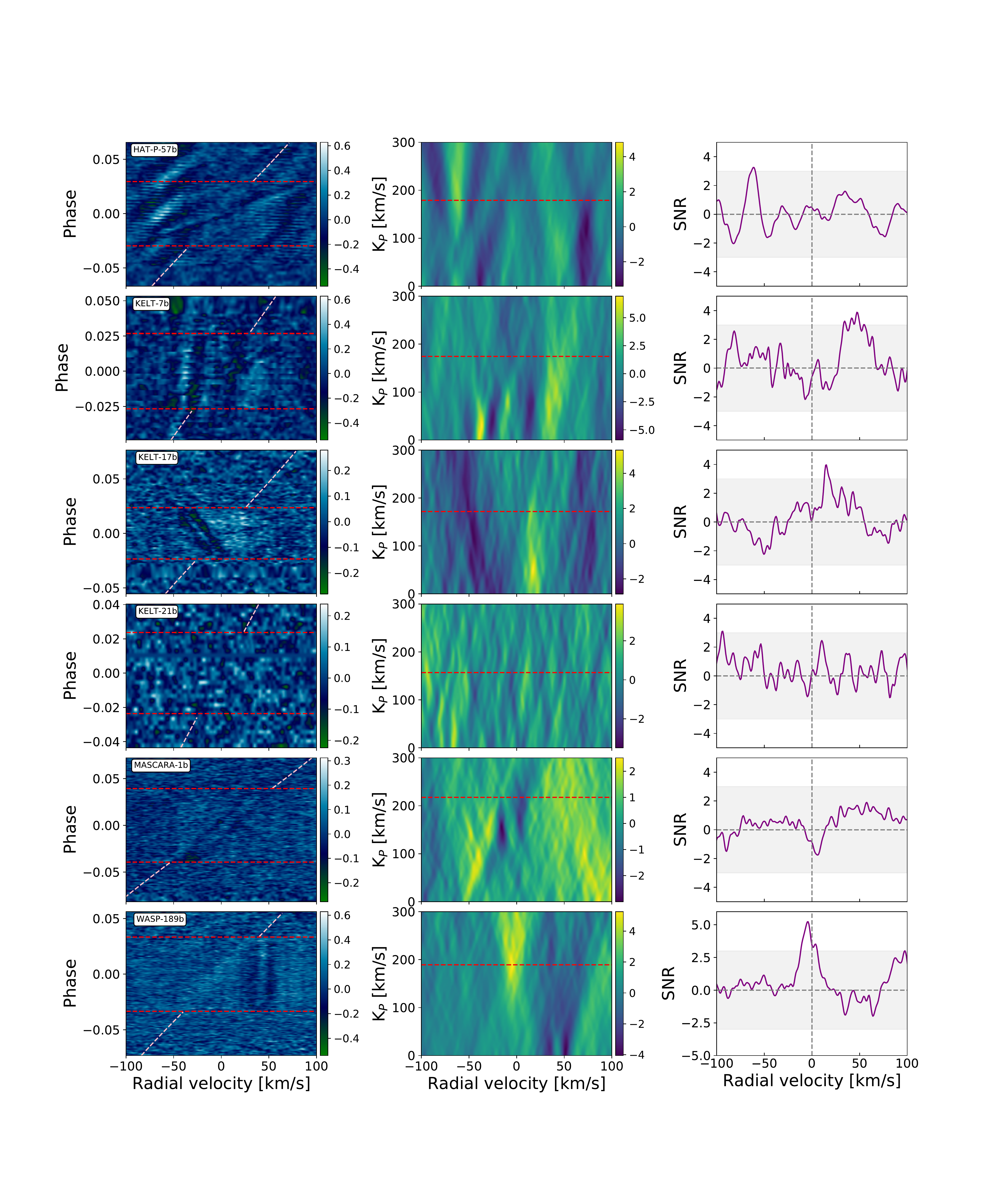}
  \caption{Cross-correlation results of \ion{Fe}{i} lines for the six UHJs, but with the RM + CLV effects correction applied. For each of the studied planets, we present the following plots. Left:\ Residuals map for \ion{Fe}{i}, where the white tilted line represents the expected velocity of the planet and red horizontal lines show the beginning and the end of the transit, region where we expect the signal from the planet; center map: $K_p$ map for $K_p$ in the range of 0 to 300~km\,s$^{-1}$ and the red line represents the theoretical $K_p$ value for the planet. We expect the signal to be coming from a planet at radial velocity of 0~km\,s$^{-1}$ and theoretical $K_p$. Right: S/N plot for expected $K_p$ value, the grey vertical lines represent 0~km\,s$^{-1}$ and gray region represents S/N between -3 and 3. In the residual maps for HAT-P-57b and KELT-7b, the remaining signal (dark and light tilted lines) comes from stellar pulsations (see Section \ref{sec:Results}).}
  \label{fig:FeI-all}
\end{figure*}

\subsection{Cross-correlation analysis}\label{sec:CC}

Using the cross-correlation (CC) method \citep{Snellen2010}, we investigated the chemical atmospheric composition of all six planets in our sample. Our analysis specifically searched for \ion{Al}{i}, \ion{B}{i}, \ion{Ca}{i}, \ion{Cr}{i}, \ion{Fe}{i}, \ion{Fe}{ii}, H$_2$O, \ion{K}{i}, \ion{Li}{i}, \ion{Mg}{i}, \ion{Mg}{ii}, \ion{N}{i}, \ion{Na}{i}, \ion{Si}{i}, \ion{Si}{ii}, \ion{Ti}{i}, TiO, \ion{V}{i}, \ion{V}{ii}, VO, and \ion{Y}{i}. Models of all species were calculated using the petitRADTRANS code \citep{molliere_2019_petiRADTRANS}, where the atomic abundances were calculated assuming solar abundances from \citet{Asplund2009SunAbund}. The line lists used for each of the species are listed in Table \ref{table:line_list}. For TiO, two different line lists were used. We emphasize that the accuracy of the line list is crucial to avoid biasing or hampering possible detections.

In addition, in the calculations, we added a cloud layer at $P_0 =1$mbar to simulate the continuum opacity produced by H$^{-}$ \citep{Hoeijmakers_2019_kelt9}.
We also considered isothermal pressure-temperature profile at $4000$K.
As a last step, we convolved the transmission model spectra with the instrumental profile of HARPS-N and HARPS.

As described in previous section,
we corrected telluric contamination using ESO tool {\tt Molecfit} \citep{Molecfit1, Molecfit2}. Next, the spectrum was divided into 
  10000-pixel-wide components (hereafter "orders") and we fitted a quadratic polynomial in time at each wavelength. Pixels that deviated from the fitted polynomial by more than $5\sigma$ were excluded. We then normalized each spectral order  by fitting a quadratic polynomial to the pseudo-continuum, and
emission lines greater than 5\% of the flux continuum were masked. To remove the stellar signal, we divided each spectrum by the master-out spectrum, the mean of all the out-of-transit spectra. 

Next, we cross-correlated the residuals in the Earth's rest frame with an atomic and molecular model over a radial velocity range of $\pm200~{\rm  km~s^{-1}}$ in steps of 0.8~km\,s$^{-1}$. Afterwards, the cross-correlation map was shifted to the planet rest frame using the formula:

\begin{equation}
    v_p(t, K_p)= K_p \sin{2 \pi \phi (t)+v_{sys} + v_{bar}(t)}
,\end{equation}

where  $v_p$ is the planet radial velocity, $K_p$ is the semi-amplitude of the exoplanet radial velocity, $\phi (t)$ is the orbital phase of a planet, $v_{sys}$ is the systemic velocity, and $v_{bar}(t)$ is the barycentric velocity. We calculated $v_p$ for a range of $K_p$ values, from 0 to 300~km\,s$^{-1}$, in steps of 1~km\,s$^{-1}$. Predicted $K_p$ values for each planet are given in Table \ref{table:param}.

Following \citet{birkby_2017_sysrem}, \citet{brogi_2018_kpmap}, \citet{alonso_floriano_2018_hd189733}, and \citet{sanchez_lopez_2019_hd209458}, the in-transit cross-correlation values (excluding ingress and egress) for each $K_p$  were co-added independently. We checked the significance of the signal by calculating its S/N by  dividing the result by the standard deviation calculated far from the position where we expected the signal (-150~km\,s$^{-1}$ to -50 ~km\,s$^{-1}$ and 50 ~km\,s$^{-1}$ to 150~km\,s$^{-1}$).

We removed the RM and CLV effects from the cross-correlation maps of different species where RM is clearly visible (i.e., \ion{Ca}{i}, \ion{Cr}{i}, \ion{Fe}{i}, \ion{Fe}{ii}, \ion{Mg}{i}, \ion{Mg}{ii}, and \ion{Si}{ii}), using the models described in Section \ref{sec:TS}. Because of the uncertainties of the parameters used to calculate the models and the high quality of some observations, we were able to detect small differences between the observed and calculated slope of the RM. In order to improve the correction we shifted and scaled the models to match the observations. 

At the S/N achieved during the observations, we note that the CLV effect for single line transmission spectroscopy is negligible, and thus we did not take it into account; however, this is not a case where thousands of lines would contribute to the CC.

Finally, in order to obtain a higher S/N and smaller uncertainties in the $K_p$ determination, when multiple transit observations were available, we combined the nights by co-adding the in-transit CC residuals for each $K_p$ value.

\section{Results}\label{sec:Results} 

\subsection{Transmission a pectroscopy of single lines}\label{sec:resTS} 

In Figure~\ref{fig:Ha-all}, we show the transmission spectrum around the H$\alpha$ line for the different planets. We did not detect any significant H$\alpha$ absorption signal for any of the planets -- except KELT-17b and WASP-189b.

For KELT-21b and HAT-P-57b, we observe that the combined RM effect in the planet rest frame is too faint to be distinguished in the data. For KELT-17b and KELT-7b, it is tentatively observed, while for WASP-189b and MASCARA-1b, the RM can be clearly identified in the final transmission spectrum thanks to the high S/N achieved. 

For WASP-189b, which presents the highest S/N (the host's magnitude is V=6.6), we observe a deviation between observations and modeled RM at the H$\alpha$ position, which has also been observed by \citet{Cauley2020_W189}. When removing the model, this deviation remains as an absorption feature of $0.13\pm0.02\%$ and FWHM of $0.6\pm0.1~{\rm \AA}$ centred at H$\alpha$, which corresponds to a  detection of $\sim6\sigma$. For MASCARA-1b, we observe that the RM models and the observations do not fully agree, obtaining significant residuals after the model subtraction. This is probably due to the low precision of the system parameters found in the literature. For this planet, due to low S/N, we discarded the observations obtained during the night of 2019-09-05 (Night 4) from our H$\alpha$ analysis (see Figure~\ref{fig:SNR}). The transmission spectrum of HAT-P-57b shows a narrow absorption residual emerging from the noise in the H$\alpha$ position. This absorption signal can be described with a Gaussian profile of $0.7\pm0.2~\%$ ($\sim3.5\sigma$) contrast and a FWHM of $\sim 0.19~{\rm \AA}$. However, stellar pulsation residuals are clearly observed in the cross-correlation analysis (e.g., Figure~\ref{fig:FeI-all_RM}), which are also present in the H$\alpha$ surroundings, although at lower S/N -- they are probably affecting our results. In addition, the absorption signal is relatively narrow when comparing to the H$\alpha$ absorption lines found in other UHJs (\citealt{seidel-2019-wasp-76, YanKELT9}, among others). Finally, for KELT-17b, the transmission spectrum shows an absorption signal of $0.8\pm0.2~\%$ ($\sim4\sigma$), FWHM$=0.4\pm0.1~{\rm \AA}$, which is significantly blue-shifted ($-7$\,km\,s$^{-1}$) with respect to the expected H$\alpha$ position. In the 2D maps (shown in Figure~\ref{fig:Ha-all}), we observe that this absorption mainly comes from the second half of the transit, and could be due to some variations in the stellar lines core. For this particular planet, more data is needed to confirm the tentative detection reported here. None of the other two UHJs studied here, namely, KELT-21b and KELT-7b, show any significant structure in the transmission spectrum around H$\alpha$.

\begin{figure}[h]
  \includegraphics[width=0.45\textwidth]{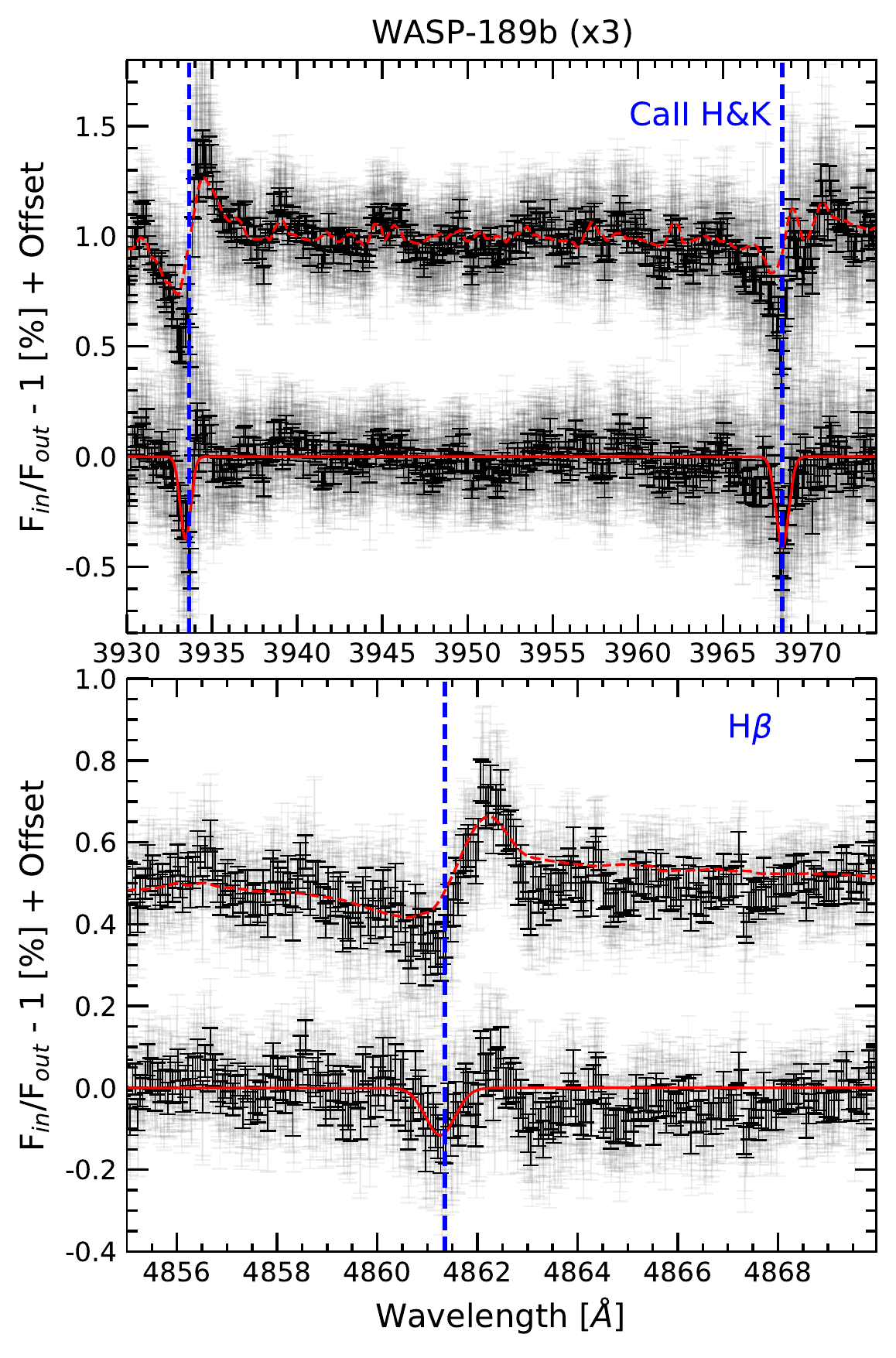}
  \caption{ Transmission spectra of WASP-189b around  \ion{Ca}{ii} H\&K (top panel) and H$\beta$ (bottom panel) lines. In the top panel, the transmission spectrum shown in black is binned by $0.1$\,{\AA}, while the transmission spectrum presented in the bottom panel is binned by $0.05$\,{\AA}. The red dashed lines show the predicted RM effect in the transmission spectrum. The red solid lines show the best-fit Gaussian profile to the transmission spectrum after correcting the RM effect. }
  \label{fig:CaHK_Hb}
\end{figure}

As for the H$\alpha$ line, we explored the full wavelength range covered by HARPS and HARPS-N observations, in pursuit of planetary absorption signals, but we did not find any significant signal in any of the six UHJs. For WASP-189b, the transmission spectrum shows strong RM features in the stellar lines position. As observed for H$\alpha$, the transmission spectrum around the strongest stellar lines experiences a similar deviation from the predicted RM model, where the absorption from the exoplanet atmosphere is expected (e.g., see Figure~\ref{fig:CaHK_Hb}). WASP-189 has shown clear evidence of gravity darkening using CHEOPS observations \citep{2020_Lendl_wasp189}. The presence of gravity darkening is expected to modify the RM effect and could produce variations as those observed in Figures~\ref{fig:Ha-all} and \ref{fig:CaHK_Hb}. On the other hand, these lines (H$\alpha$, H$\beta$, \ion{Na}{i}, \ion{Ca}{ii}, etc.) are expected in the transmission spectra of UHJ (e.g., \citealt{seidel-2019-wasp-76,Casasayas2018}), which complicates the interpretation of our results. Thus, further modeling efforts that include the effects of gravity darkening are needed to clearly conclude whether the deviations observed between the models and the observations are due to the particular atmosphere of WASP-189b.

\subsection{Cross-correlation analysis} 

Figures~\ref{fig:FeI-all_RM} and \ref{fig:FeI-all} show the results of the cross-correlation analysis for each planet, aimed at testing for the presence of \ion{Fe}{i}. 
In the right column, the cross-correlation residuals for each of the studied planets are shown, after combining all available nights; the middle columns show the significance maps for $K_p$ in the range of 0 to 300~km\,s$^{-1}$, where red horizontal line indicates the expected $K_p$ value; the right columns present the S/N plot at the expected $K_p$ value. Figure~\ref{fig:FeI-all_RM} presents the results without correction for the RM+CLV effects, while Figure~\ref{fig:FeI-all} shows the results after our correction. 

In the Appendix, similar figures presenting the cross-correlation residual maps and S/N plots are given for each of the atomic and molecular species we explored, namely: In Fig. \ref{fig:Al-all} (\ion{Al}{i}), \ref{fig:B-all} (\ion{B}{i}), \ref{fig:CaI-all} (\ion{Ca}{i}), \ref{fig:Cr-all} (\ion{Cr}{i}), \ref{fig:FeI-all} (\ion{Fe}{i}), \ref{fig:FeII-all} (\ion{Fe}{ii}), \ref{fig:H2O-all} (H$_2$O), \ref{fig:K-all} (\ion{K}{i}), \ref{fig:Li-all} (\ion{Li}{i}), \ref{fig:MgI-all} (\ion{Mg}{i}), \ref{fig:MgII-all} (\ion{Mg}{ii}), \ref{fig:Na-all} (\ion{Na}{i}), \ref{fig:Si-all} (\ion{Si}{i}), \ref{fig:SiII-all} (\ion{Si}{ii}), \ref{fig:Ti-all} (\ion{Ti}{i}), \ref{fig:TiO-all} (TiO), \ref{fig:V-all} (\ion{V}{i}), \ref{fig:VII-all} (\ion{V}{ii}), \ref{fig:VO-all} (VO), and \ref{fig:Y-all} (\ion{Y}{i}).

\begin{figure*}[h]
  \includegraphics[width=0.9\textwidth]{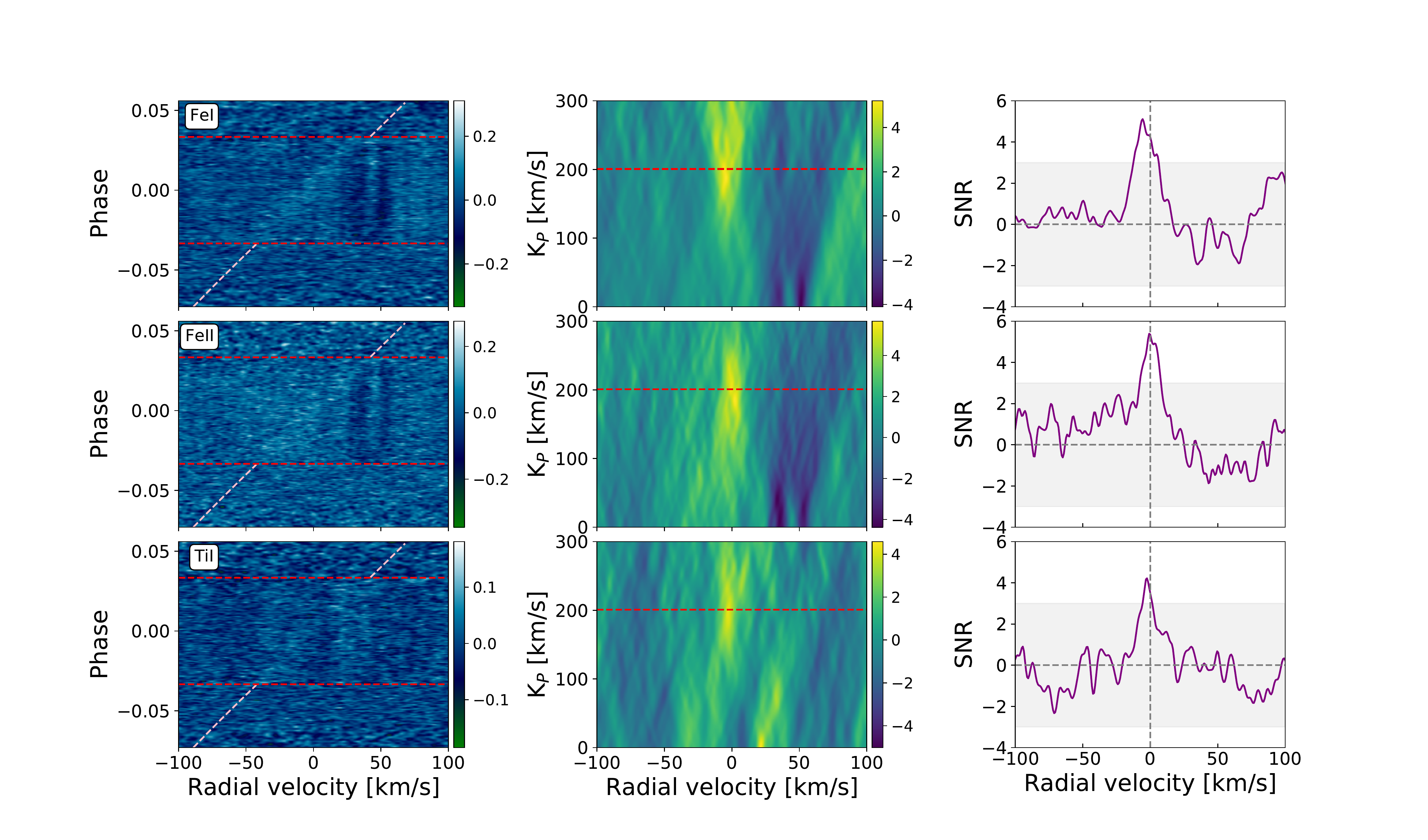}
\caption{Detections of \ion{Fe}{i}, \ion{Fe}{ii,} and \ion{Ti}{i} in the atmosphere of WASP-189b. The planetary signal appears at the expected $K_p$ value.}
  \label{fig:W189_detections}
\end{figure*}

Overall, we did not detect any significant atmospheric signal for five of the six studied planets, with the exception again of WASP-189b, for which we detected the planetary absorption features of \ion{Fe}{i}, \ion{Fe}{ii,} and \ion{Ti}{i}, (cf. Figure \ref{fig:W189_detections}). The detection of \ion{Fe}{i} confirms previous studies carried out by \cite{Yan_2020_wasp-189}, demonstrating the detection of \ion{Fe}{i} in the dayside of the planet using emission spectroscopy. \ion{Fe}{ii} and \ion{Ti}{i} were detected for the first time in the atmosphere of this planet. We detected the signal consistent with the exoplanet radial velocity change at 5.1 $\sigma$ for \ion{Fe}{i}, 5.4 $\sigma$ for \ion{Fe}{ii}, and 4.2 $\sigma$ for \ion{Ti}{i}. In Figure \ref{fig:SNR_N3_S3}, we present the signal coming from separated nights and it is clear that it was necessary to combine more than one night in order to achieve a signal with a S/N higher than 3.

HAT-P-57 is a pulsating star \citep{Hartman_hat-p-57}, thus, it exhibits strong pulsation signals in CC maps, especially for \ion{Ca}{i}, \ion{Fe}{i}, and \ion{Fe}{ii}, which are visible in the two individual transit observations. This signal can block or mimic the planetary atmospheric signal and it is extremely difficult to disentangle them. For KELT-7b, we detected a strong vertical signal (on both of the nights), which can also be related to stellar pulsations, which were observed in the Doppler tomographic analysis of lower resolution data from FIES \citep{Kelt-7_pulsations}.

Studies of the atmosphere of MASCARA-1b are challenging due to the impact of the RM effect, which shows the same radial-velocity change as the expected signal coming from the planet, which is strongly visible in the CC of \ion{Fe}{i} (see Figure \ref{fig:FeI-all_RM}). For atoms where RM is detectable (\ion{Ca}{i}, \ion{Cr}{i}, \ion{Fe}{i}, \ion{Fe}{ii}, \ion{Mg}{i}, \ion{Mg}{ii}, and \ion{Si}{ii}), any possible planetary signal can be masked by this effect. For species that are not present in the stellar spectrum (hence, having no RM signal), we did not detect any signal.

\section{Discussion and conclusions}\label{sec:Discussion} 

\begin{figure*}[h]
  \includegraphics[width=0.9\textwidth]{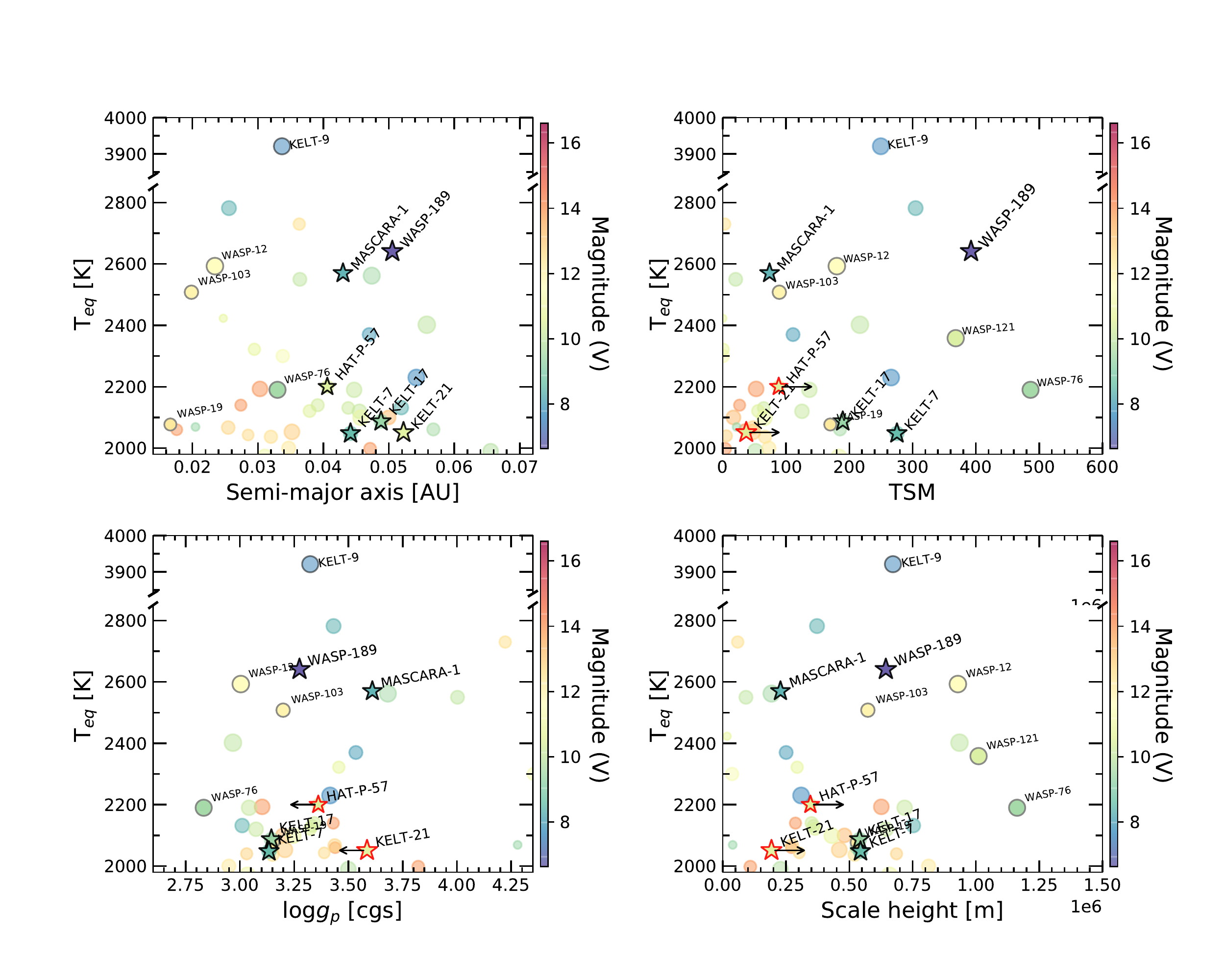}
  \caption{Equilibrium temperature versus several planet parameters for the known ultra-hot Jupiter (Teq > 2000 K) planet population. Planets studied in this work are marked with a star symbol. Top left:\ Planet-to-star distance (semi major axis) is shown in the horizontal axis and the equilibrium temperature of the exoplanets is shown in the vertical axis. We note that only those planets with Rp > 0.6 RJ are shown. Bottom left: Context of studied planets, marked with a star symbol, with respect to all the UHJ with known surface gravity (we excluded all planets whose masses have only upper limits). The equilibrium temperature is shown in the vertical axis and surface gravity of the planets in the horizontal axis. Top right: Calculated transmission spectroscopy metrics (TSM) versus equilibrium temperature of the exoplanets, where stars represent the planets studied in this work. Bottom left: Calculated scale height versus equilibrium temperature of the exoplanets. 
  For all of the figures, the V band magnitude of the host star is color-coded and the marker size is indicative of the planetary radius (data are extracted from exoplanet.eu). }
  \label{fig:logg}
\end{figure*}

In Figure \ref{fig:logg}, we present several plots that include all known UHJs,  where the equilibrium temperatures of the planets are plotted against semi-major axis, transmission spectroscopy metrics \citep{Kempton2018}, $\log g$, and atmospheric scale height ($H$). The color of the symbols represent the brightness of the host star. A priori the sample of UHJs studied here cover a broad parameter space and atmospheric characterization of the full sample  could have yield several atomic species. However, as discussed in previous sections, except for WASP-189b, we did not detect any atmospheric species. These non-detections can be explained by way of a few possible scenarios.

\begin{figure*}[h]
  \includegraphics[width=0.9\textwidth]{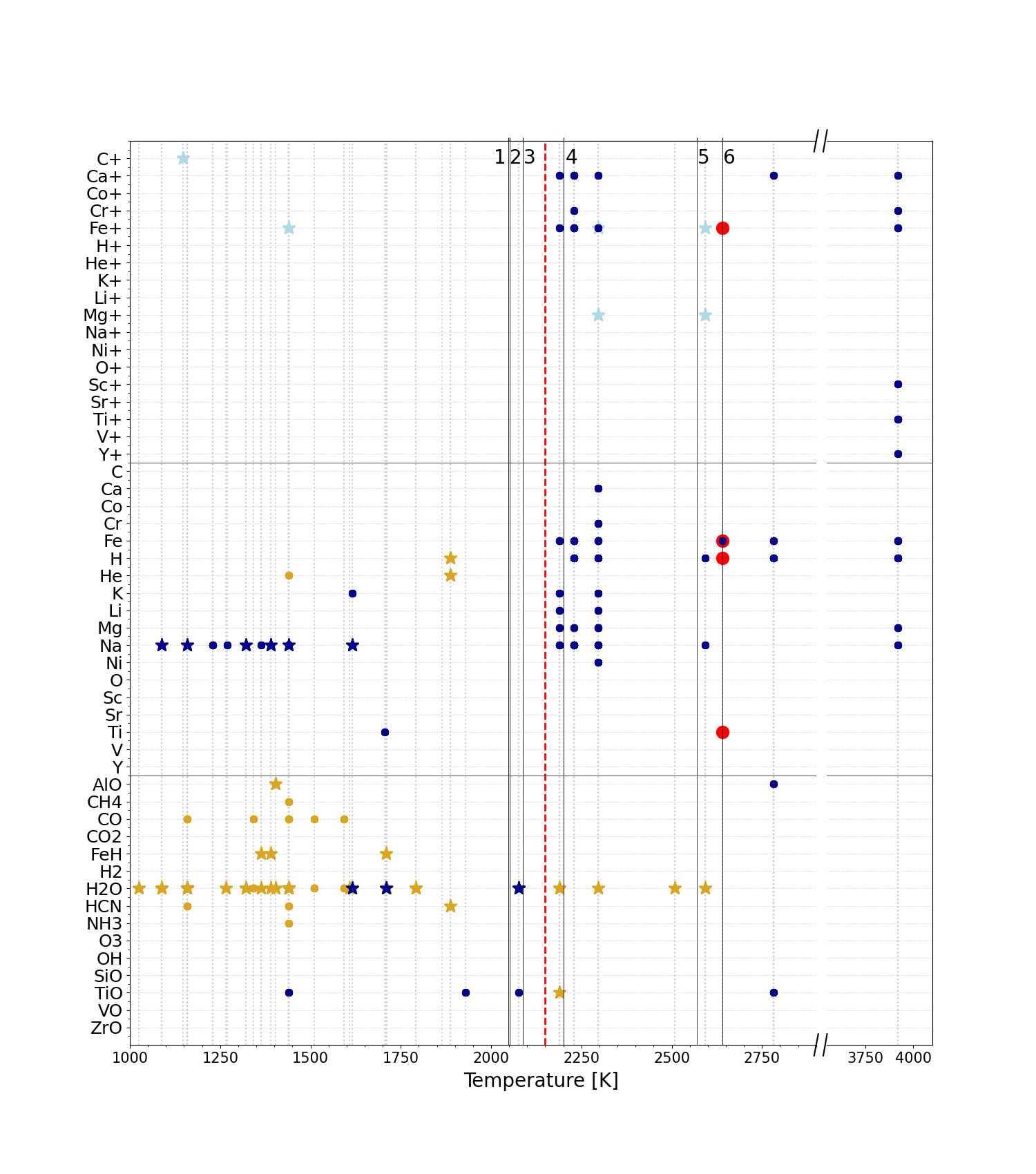}
\caption{Plot of all detected species in planets with $T_{eq} > 1000 K$, whose atmospheres (day- and nightside) have been studied using ground- and space-based telescopes. The atoms, ions, and molecules are shown in the vertical axis against planet temperature. Circles represent species detected using ground-based telescopes, and stars using space-based telescopes. Light-blue symbols show detections using UV observations, dark-blue symbols correspond to visible observations, and orange symbols correspond to near-IR observations. Each vertical grey line represents a planet whose atmosphere has been explored. The dashed red line marks $T_{eq} = 2150 K$. The positions of the 6 UHJs studied in this paper are labeled and marked with solid vertical lines -- 1: KELT-7b, 2: KELT-21b, 3: KELT-17b, 4: HAT-P-57b, 5: MASCARA-1b, 6: WASP-189b). Finally, the red circles indicate the new detections reported in this work. Data are taken from the \texttt{ExoAtmospheres} database.} 
  \label{fig:ExoAtmo}
\end{figure*}

First and foremost, the non-detections could respond to the absence of the studied species in the atmospheres of these planets. Three of our planets are "cold" UHJs in the temperature range of 2000-2200 K,  where ionized metals have not yet been detected (see Figure \ref{fig:ExoAtmo}). HAT-P-57b lies just at the 2200 K frontier, but MASCARA-1b and WASP-189b are among the hottest known UHJs.

Another factor is that half of these planets are not favorable targets for transmission spectroscopy, due to a high surface gravity, low brightness of the host star, or a combination of both. Their transmission spectroscopy metrics (TSM, \citet{Kempton2018}) are: KELT-7 b (275.21), KELT-17 b (189.64), KELT-21 b (37.11), HAT-P-57 b (88.45), WASP-189 b (392.29), and MASCARA-1 b (73.95). As indicated by \citet{Kempton2018}, giant planets with TSM greater than 90 are targets for high-quality atmospheric characterization. Three planets in our sample have TSM smaller than 90, which could explain the non-detection in their atmospheres. For comparison, the TSM for well-studied UHJs and HJs are: KELT-9b (249.74), MASCARA-2b (266.04), WASP-12 b (180.26), WASP-33 b (304.69), WASP-76 b (486.31), HD 209458b (953.62),  WASP-69b (835.65), and HAT-P-32b (331.78). Additionally, for KELT-21b, only one transit observation is available and the relatively low S/N could also contribute to the lack of atmospheric detections.

The presence of a strong RM effect also complicates the detection of a planetary atmosphere and is particularly poignant for our sample. This effect is clearly visible in the residual maps for all six planets. The RM effect is very important for MASCARA-1b, as its trace lies directly on top of any possible planetary signal, making it impossible to disentangle the two signals, even after trying to correct for this effect (see Figure \ref{fig:FeI-all}). As we can see in Figures \ref{fig:FeI-all_RM} and \ref{fig:FeI-all}, even for WASP-189b, it is necessary to correct the final CC maps for a strong RM effect in order to be able to detect the planetary signals. Nonetheless, some residual structures, resulting from a non-perfect correction, are visible in the residual maps. For KELT-7b, the RM also partly overlaps with the planetary trail. 

Finally, in the case of HAT-P-57b and KELT-7b, the effects of stellar pulsations are also visible in the CC residual maps and those can cover or mimic the planetary signal.

In summary, while half of our targets are potentially good targets for the detection of atmospheric species (except for WASP-189b), all suffer from different effects that impede these detections. The most favorable candidate for transmission spectroscopy, KELT-7b, presents strong stellar pulsations in their CC residual maps that can cover or mimic the planetary signal. The other targets are affected by strong RM, especially MASCARA-1b. In addition, KELT-21b has only one transit observation that is, albeit, of a relatively "low" equilibrium temperature and TSM.

To set our findings in a broader context, in Figure \ref{fig:ExoAtmo}, we compile all the atmospheric detections for hot and ultra-hot planets, studied using ground- and space-based telescopes. Data are taken from the \texttt{ExoAtmospheres} database\footnote{http://research.iac.es/proyecto/exoatmospheres/index.php}. In Figure \ref{fig:ExoAtmo}, we present the most common atoms and molecules that can be detected in the atmospheres of giant planets, together with the current detections obtained with space-based and ground-based telescopes.

Theoretical calculations \citep{Parmentier2018, helling_wasp_18} show that in the UHJ regime, atoms in their neutral and ionised states are expected to be very abundant. Molecules such as H$_2$O may undergo dissociation on the dayside and recombination on the nightside \citep{Arcangeli2018,Gandhi2020}, enhancing, hence, the mixing ratios of ions at the expense of molecules. Metal oxides, such as CO, TiO, and VO, require much higher temperatures to be dissociated. 

While we did not detect any molecules in the sample of UHJs studied here, Figure \ref{fig:ExoAtmo} empirically illustrates how H$_2$O can indeed be detected, at least up to $T_{eq}$ of about 2600 K, as well as TiO at least up to $T_{eq}$ of about 2800 K. With the current sample of explored planets, the frontier between hot and ultra-hot atmospheres, signaled by the dominant presence of ionic species, is established at 2150 K.

%

\begin{acknowledgements}
 Based on observations made with the Italian Telescopio Nazionale Galileo (TNG) operated on the island of La Palma by the Fundación Galileo Galilei of the INAF (Istituto Nazionale di Astrofisica) at the Spanish Observatorio del Roque de los Muchachos of the Instituto de Astrofisica de Canarias.
 This work is partly financed by the Spanish Ministry of Economics and Competitiveness through grants PGC2018-098153-B-C31.
 M. S. and J. O. M. acknowledge the support of the Instituto de Astrofísica de Canarias via an Astrophysicist Resident fellowship. G. M. has received funding from the European Union's Horizon 2020 research and innovation programme under the Marie Sk\l{}odowska-Curie grant agreement No. 895525. We acknowledge funding from the European Research Council under the European Union's Horizon 2020 research and innovation program under grant agreement No 694513.

 This work made use of PyAstronomy and of the VALD database, operated at Uppsala University, the Institute of Astronomy RAS in Moscow, and the University of Vienna.

\end{acknowledgements}

\bibliographystyle{m2.bst} 
\bibliography{m2.bib} 
  

\onecolumn

\begin{appendix}
\section{S/N values}

\begin{figure*}[h]
  \includegraphics[width=\textwidth]{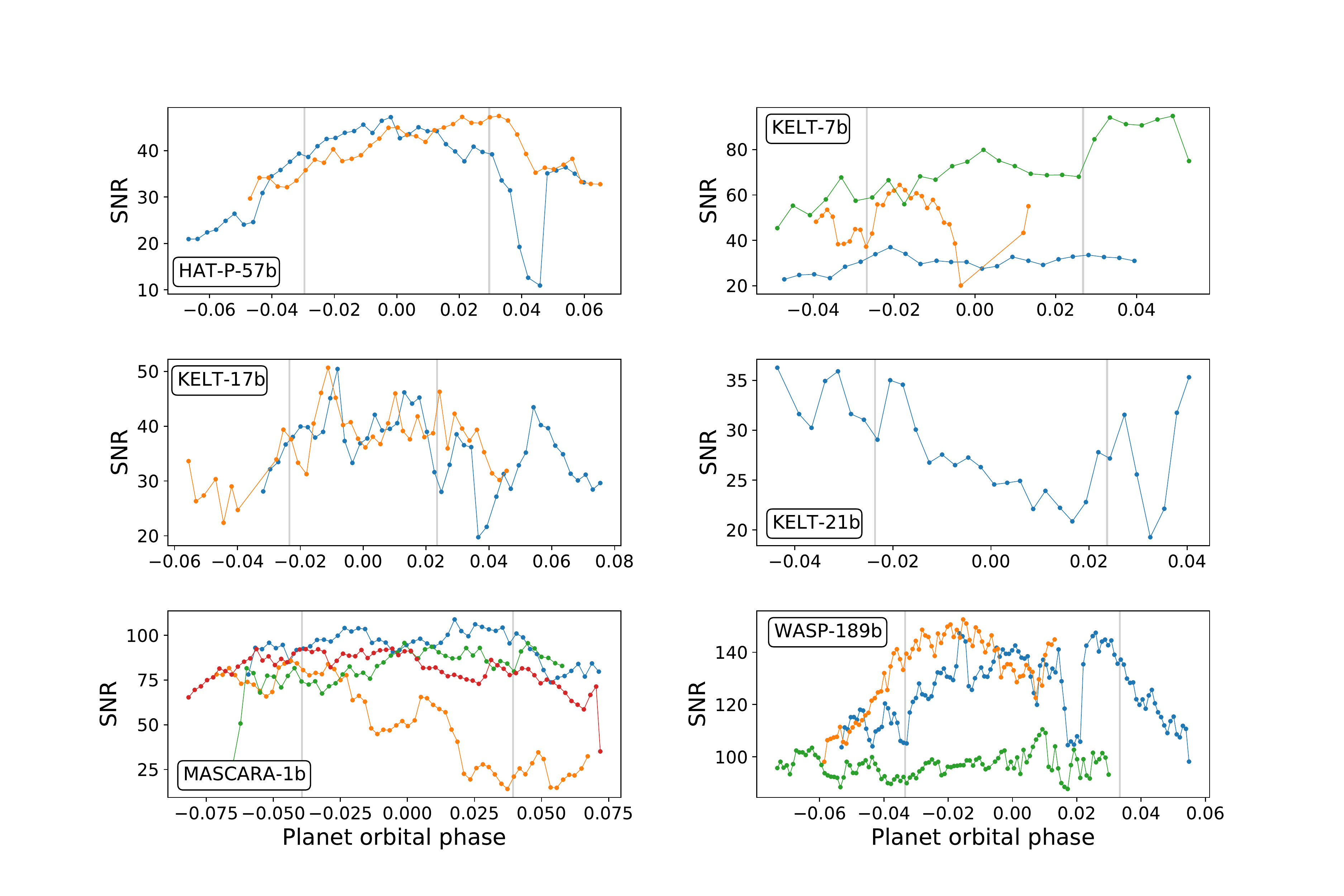}
  \caption{S/N plot during the observations. Blue plot shows the S/N of the first nights, then orange:\ second, green:\ third, and red:\ fourth. Gray vertical lines represent the beginning and end of the transit. }
  \label{fig:SNR}
\end{figure*}

\begin{table*}[h]
\caption{Line lists used in the cross-correlation analysis. }             
\label{table:line_list}      
\centering          
{\begin{tabular}{c c}     
\hline\hline       
Atom or molecule & Line list   \\
[0.2em]
\hline 

\ion{Al}{i} & Kurucz \\
\hline

\ion{B}{i} & Kurucz\\
\hline

\ion{Ca}{i} & Kurucz \\
\hline

\ion{Cr}{i} & Kurucz\\
\hline

\ion{Fe}{i} & Kurucz\\
\hline

\ion{Fe}{ii} &Kurucz\\
\hline

H$_2$O & HITEMP\\
\hline

\ion{K}{i} & VALD\\
\hline

\ion{Li}{i} &Kurucz \\
\hline
\ion{Mg}{i} &Kurucz \\
\hline

\ion{Mg}{ii} &Kurucz \\
\hline

\ion{Si}{i} & Kurucz\\

\hline

\ion{Si}{ii} &Kurucz \\

\hline

\ion{Ti}{i} &Kurucz \\

\hline

TiO & B. Plez \citep{molliere_2019_petiRADTRANS} \\
 & ExoMolOP \\
\hline

\ion{V}{i} & Kurucz\\

\hline

\ion{V}{ii} & Kurucz\\

\hline

VO & ExoMolOP\\

\hline

\ion{Y}{i} & Kurucz\\

\hline 
\end{tabular}}

\end{table*}

\begin{figure*}[h]
  \includegraphics[width=\textwidth]{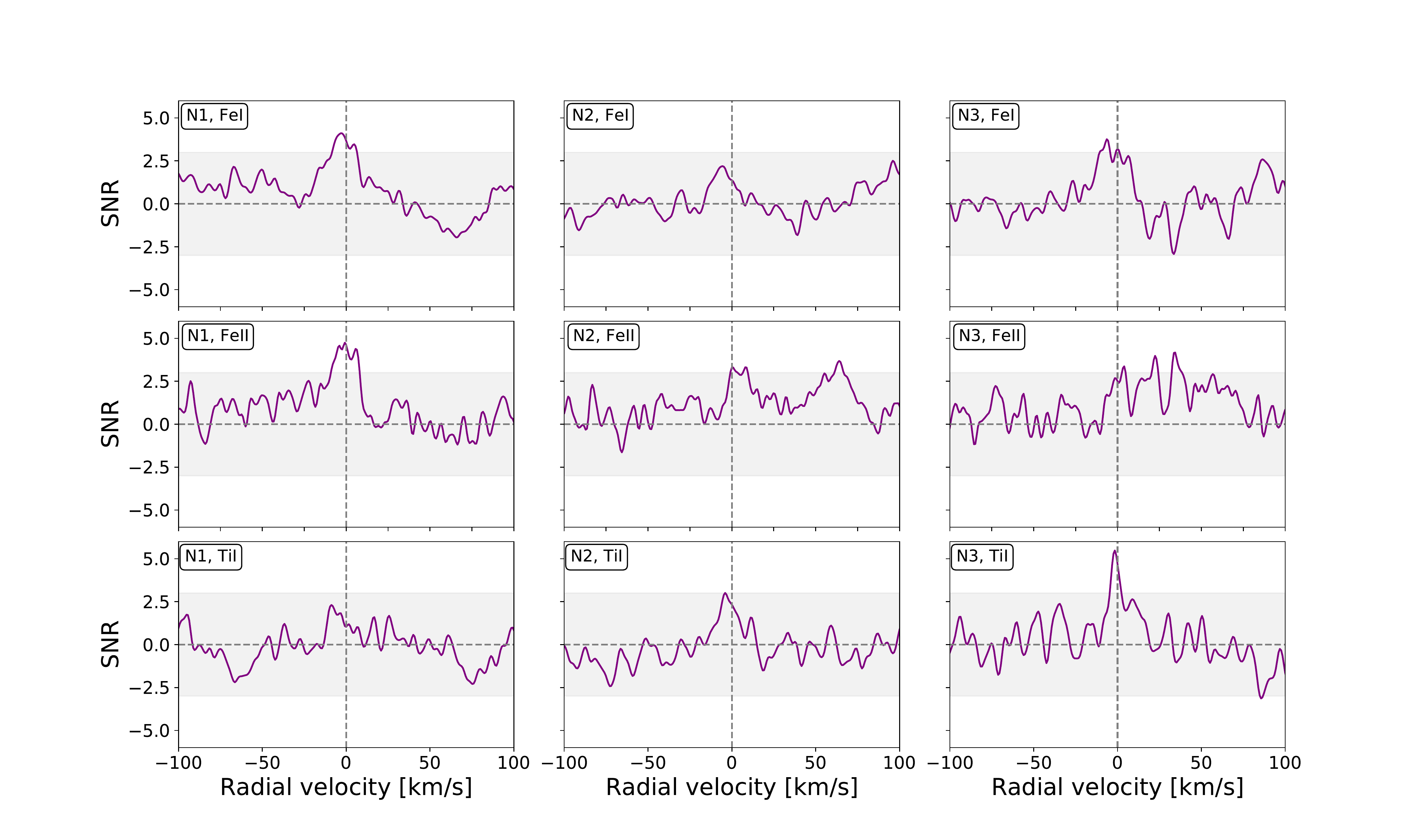}
  \caption{S/N plots for detection of \ion{Fe}{i}, \ion{Fe}{ii,} and \ion{Ti}{i} in the atmosphere of WASP-189b for each night separately. Gray regions represent S/N between -3 and 3.}
  \label{fig:SNR_N3_S3}
\end{figure*}

\begin{figure*}[h]
  \includegraphics[width=\textwidth]{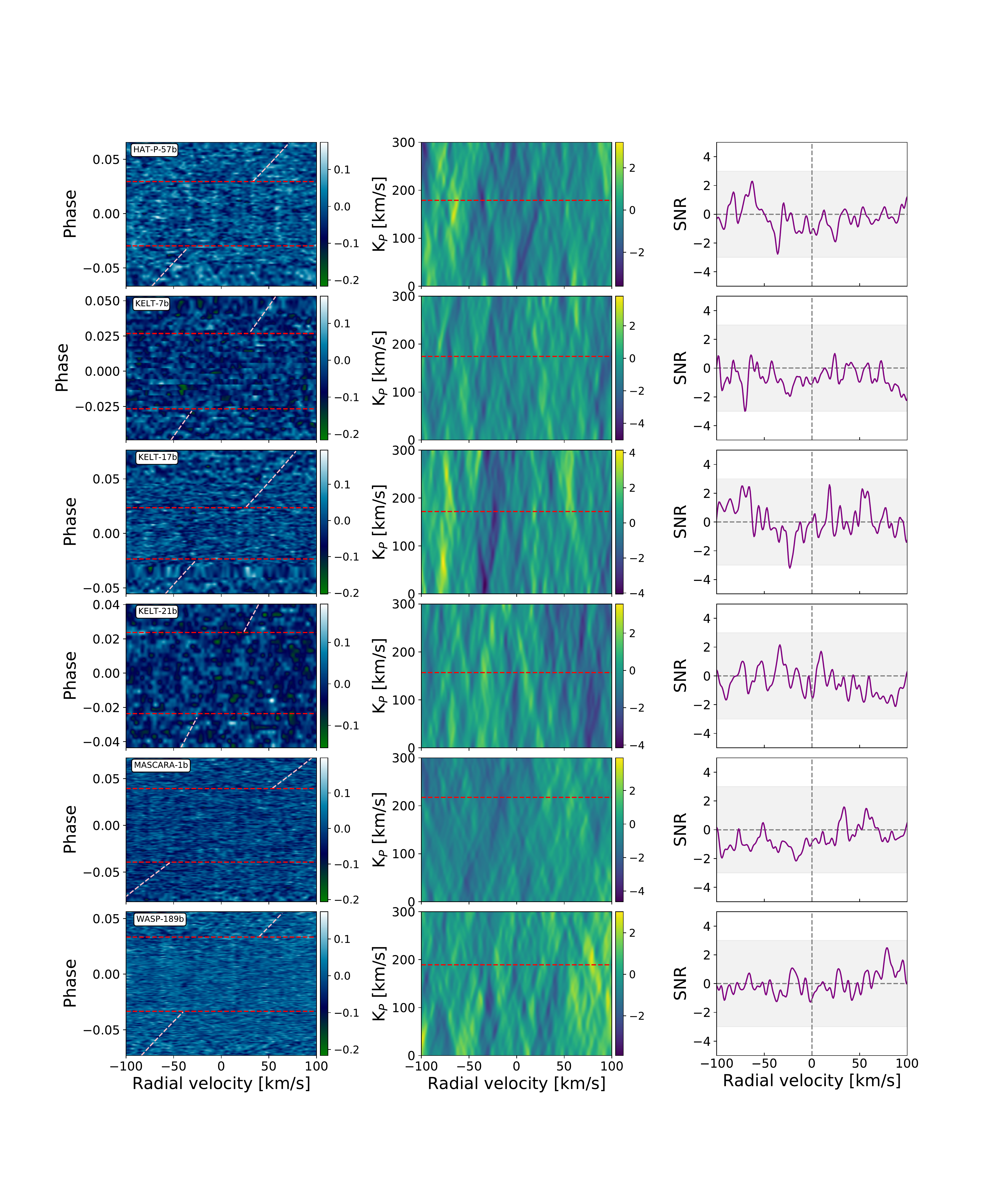}
  \caption{Cross-correlation results of \ion{Al}{i} lines for the six UHJs. For each of the studied planets, we present the following plots. Left:\ Residuals map for \ion{Al}{i}, where the white tilted line represents the expected velocity of the planet and red horizontal lines show the beginning and the end of the transit, region where we expect the signal from the planet; center map: $K_p$ map for $K_p$ in the range of 0 to 300~km\,s$^{-1}$ and the red line represents the theoretical $K_p$ value for the planet. We expect the signal to be coming from a planet at radial velocity of 0~km\,s$^{-1}$ and theoretical $K_p$. Right: S/N plot for expected $K_p$ value, the grey vertical lines represent 0~km\,s$^{-1}$ and gray region represents S/N between -3 and 3.}
  \label{fig:Al-all}
\end{figure*} 

\begin{figure*}[h]
  \includegraphics[width=\textwidth]{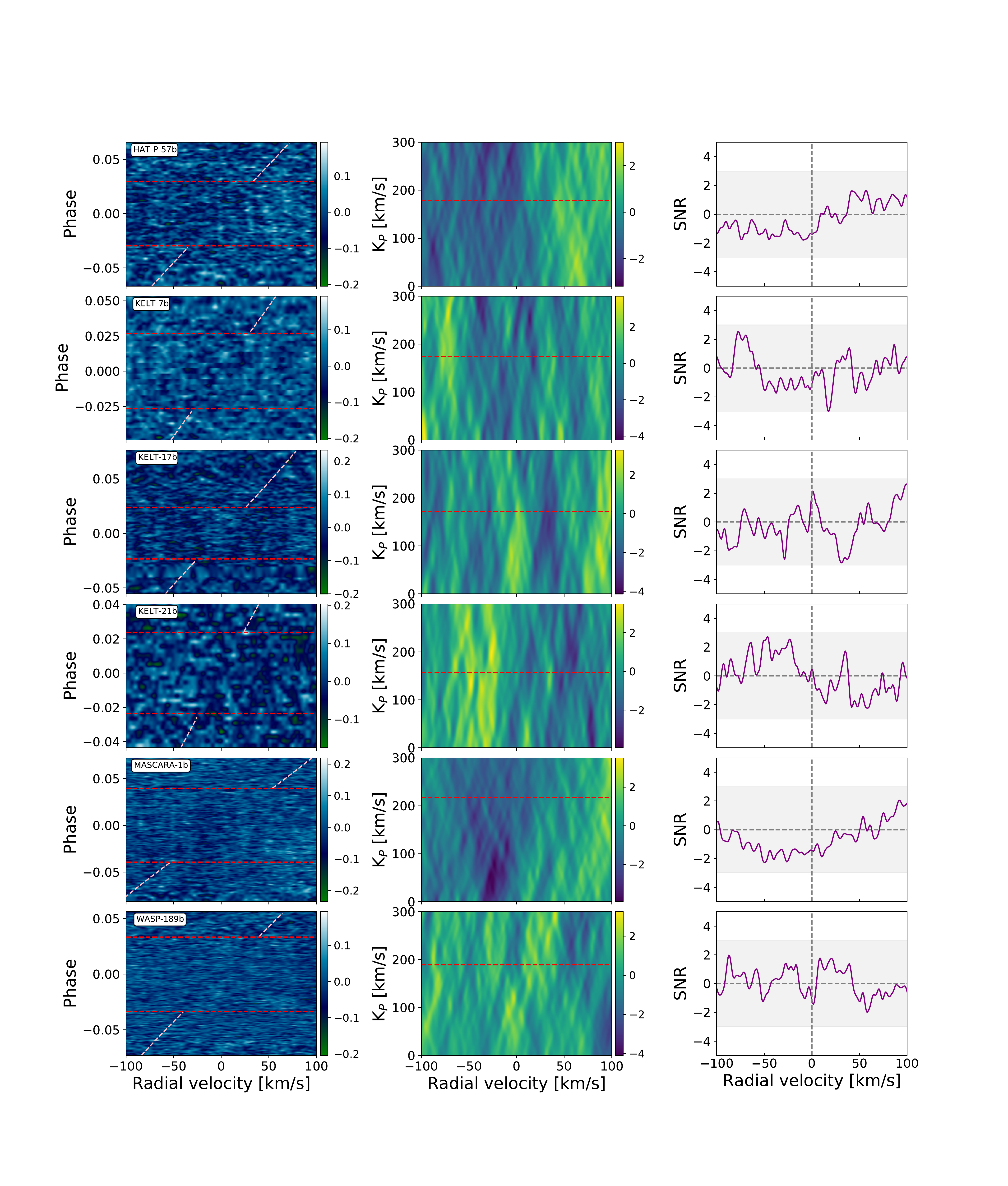}
      \caption{Cross-correlation results of \ion{B}{i} lines for the six UHJs. For each of the studied planets, we present the following plots. Left:\ Residuals map for \ion{B}{i}, where the white tilted line represents the expected velocity of the planet and red horizontal lines show the beginning and the end of the transit, region where we expect the signal from the planet; center map: $K_p$ map for $K_p$ in the range of 0 to 300~km\,s$^{-1}$ and the red line represents the theoretical $K_p$ value for the planet. We expect the signal to be coming from a planet at radial velocity of 0~km\,s$^{-1}$ and theoretical $K_p$. Right: S/N plot for expected $K_p$ value, the grey vertical lines represent 0~km\,s$^{-1}$ and gray region represents S/N between -3 and 3.}
      
  \label{fig:B-all}
\end{figure*}

\begin{figure*}[h]
  \includegraphics[width=\textwidth]{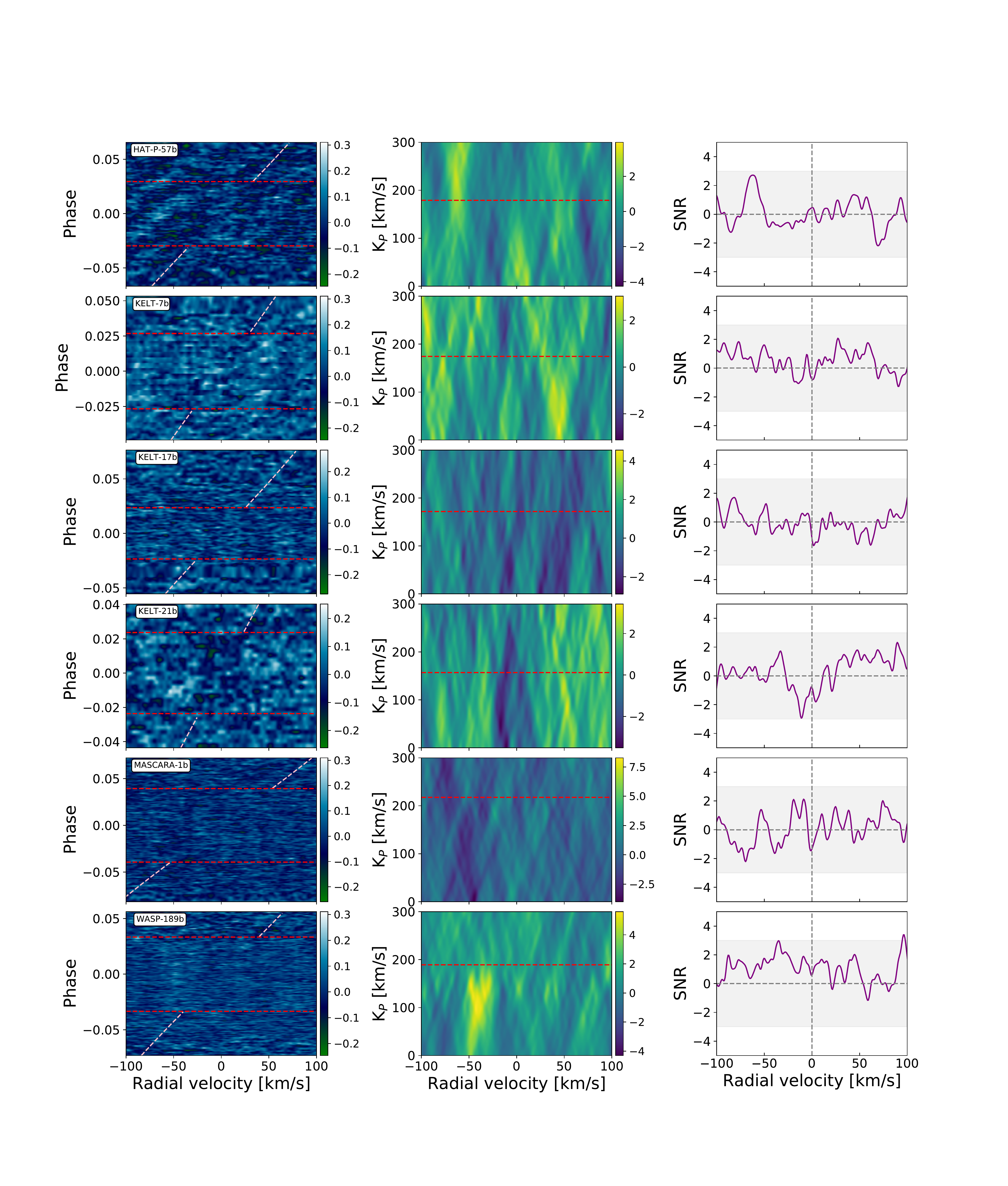}
   \caption{Cross-correlation results of \ion{Ca}{i} lines for the six UHJs. For each of the studied planets, we present the following plots. Left:\ Residuals map for \ion{Ca}{i}, where the white tilted line represents the expected velocity of the planet and red horizontal lines show the beginning and the end of the transit, region where we expect the signal from the planet; center map: $K_p$ map for $K_p$ in the range of 0 to 300~km\,s$^{-1}$ and the red line represents the theoretical $K_p$ value for the planet. We expect the signal to be coming from a planet at radial velocity of 0~km\,s$^{-1}$ and theoretical $K_p$. Right: S/N plot for expected $K_p$ value, the grey vertical lines represent 0~km\,s$^{-1}$ and gray region represents S/N between -3 and 3.}
  \label{fig:CaI-all}
\end{figure*} 

\begin{figure*}[h]
  \includegraphics[width=\textwidth]{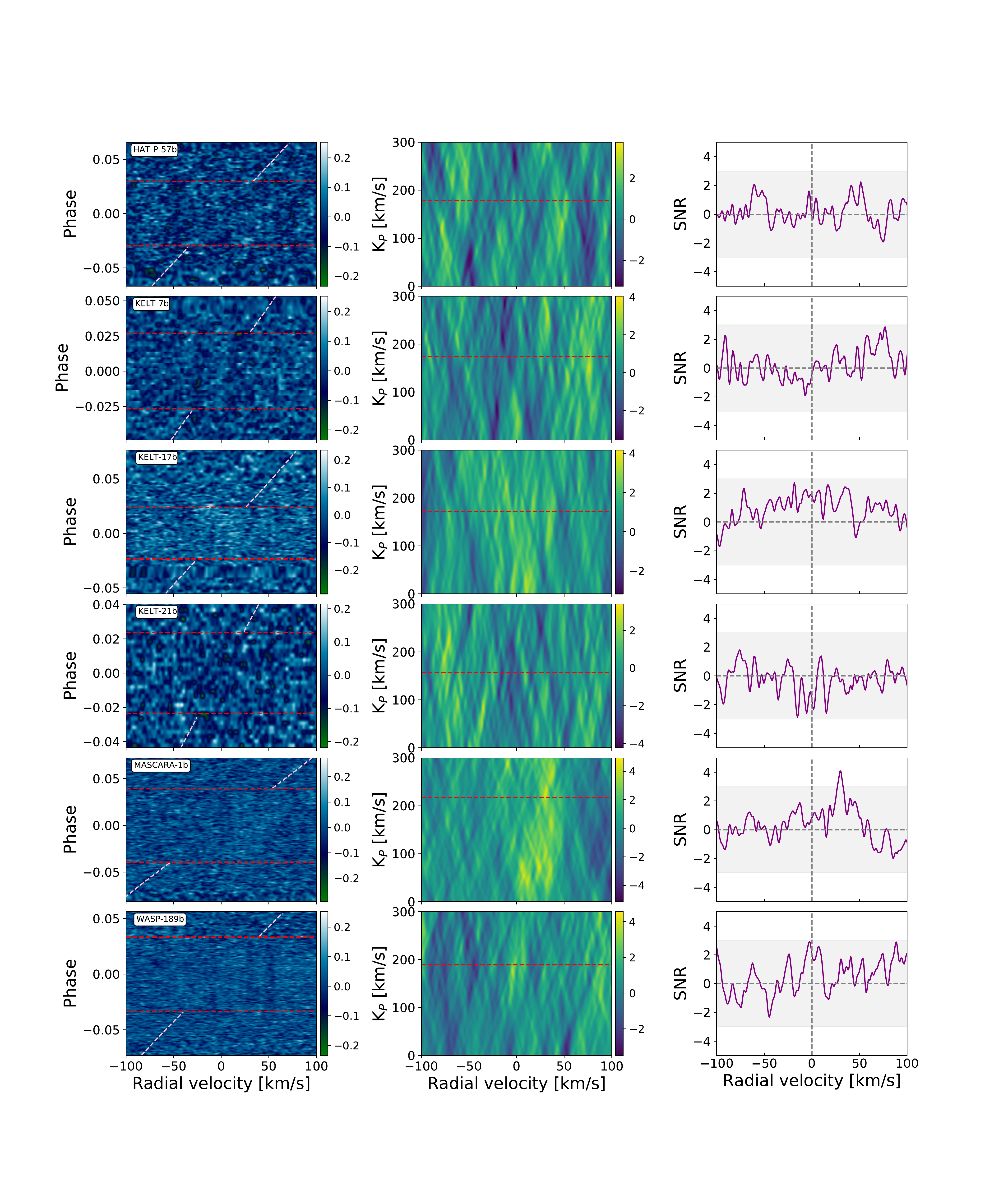}
    \caption{Cross-correlation results of \ion{Cr}{i} lines for the six UHJs. For each of the studied planets, we present the following plots. Left:\ Residuals map for \ion{Cr}{i}, where the white tilted line represents the expected velocity of the planet and red horizontal lines show the beginning and the end of the transit, region where we expect the signal from the planet; center map: $K_p$ map for $K_p$ in the range of 0 to 300~km\,s$^{-1}$ and the red line represents the theoretical $K_p$ value for the planet. We expect the signal to be coming from a planet at radial velocity of 0~km\,s$^{-1}$ and theoretical $K_p$. Right: S/N plot for expected $K_p$ value, the grey vertical lines represent 0~km\,s$^{-1}$ and gray region represents S/N between -3 and 3.}
  \label{fig:Cr-all}
\end{figure*}

\begin{figure*}[h]
  \includegraphics[width=\textwidth]{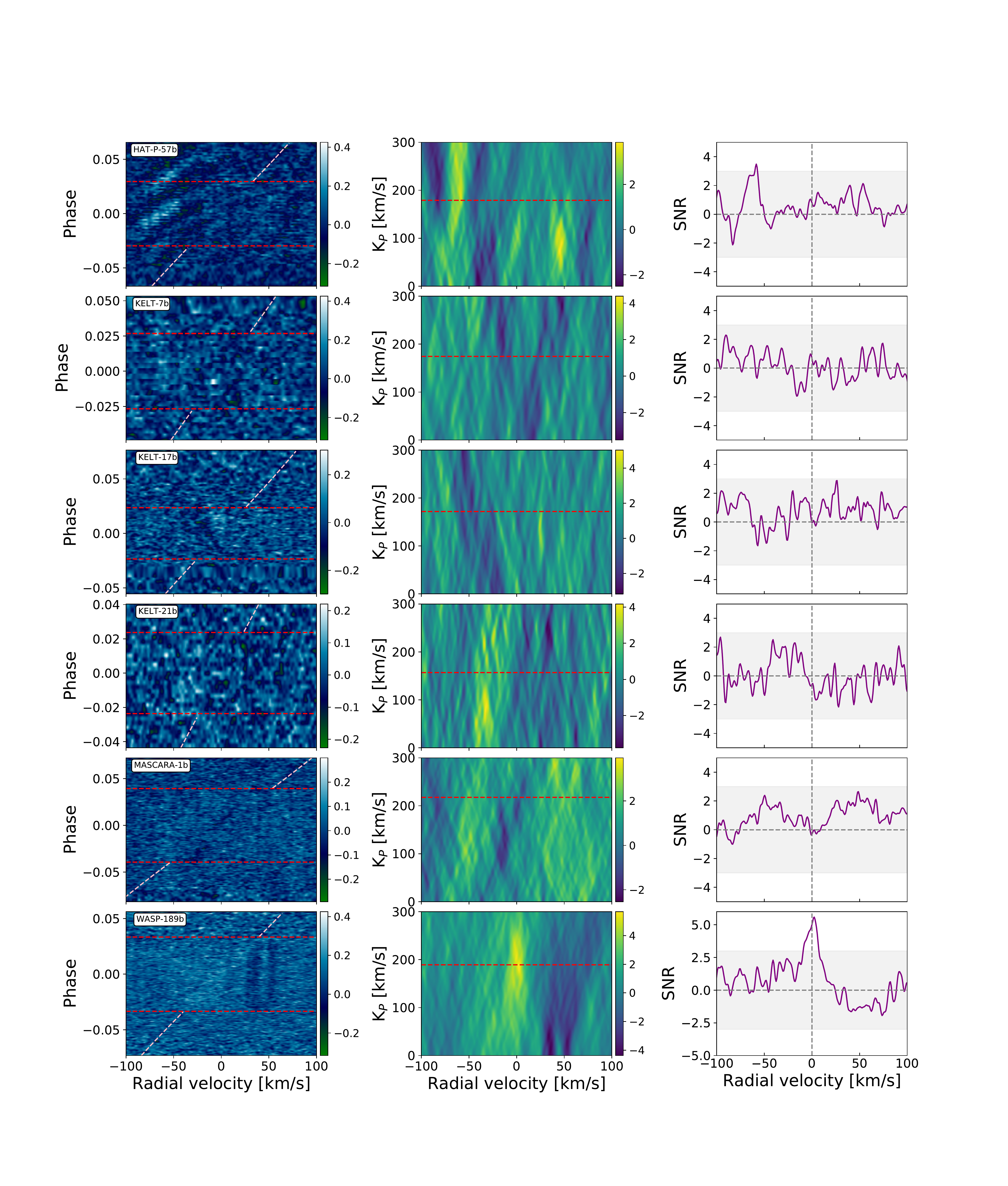}
     \caption{Cross-correlation results of \ion{Fe}{ii} lines for the six UHJs. For each of the studied planets, we present the following plots. Left:\ Residuals map for \ion{Fe}{ii}, where the white tilted line represents the expected velocity of the planet and red horizontal lines show the beginning and the end of the transit, region where we expect the signal from the planet; center map: $K_p$ map for $K_p$ in the range of 0 to 300~km\,s$^{-1}$ and the red line represents the theoretical $K_p$ value for the planet. We expect the signal to be coming from a planet at radial velocity of 0~km\,s$^{-1}$ and theoretical $K_p$. Right: S/N plot for expected $K_p$ value, the grey vertical lines represent 0~km\,s$^{-1}$ and gray region represents S/N between -3 and 3.}
  \label{fig:FeII-all}
\end{figure*}

\begin{figure*}[h]
  \includegraphics[width=\textwidth]{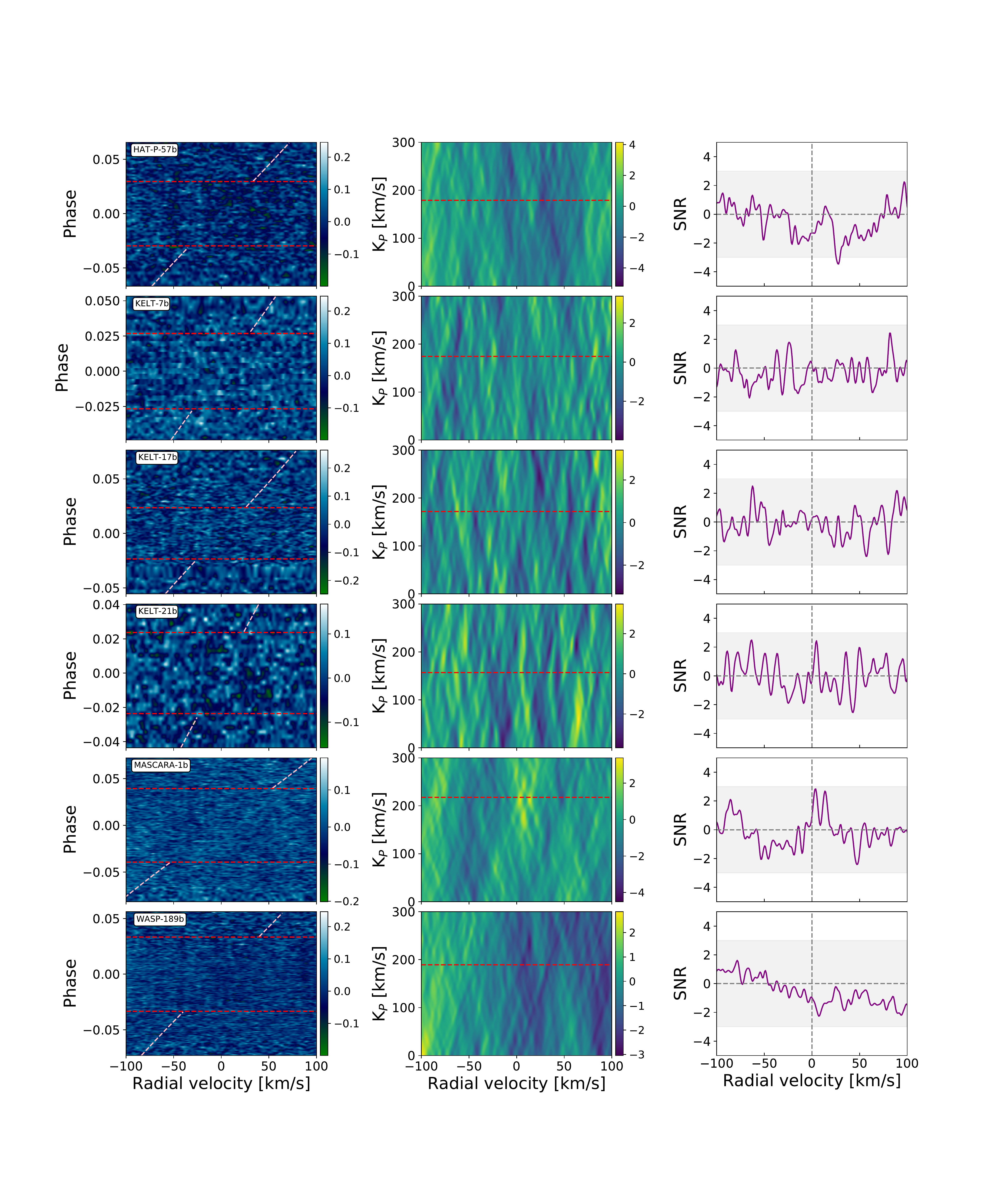}
       \caption{Cross-correlation results of H2O lines for the six UHJs. For each of the studied planets, we present the following plots. Left:\ Residuals map for H2O, where the white tilted line represents the expected velocity of the planet and red horizontal lines show the beginning and the end of the transit, region where we expect the signal from the planet; center map: $K_p$ map for $K_p$ in the range of 0 to 300~km\,s$^{-1}$ and the red line represents the theoretical $K_p$ value for the planet. We expect the signal to be coming from a planet at radial velocity of 0~km\,s$^{-1}$ and theoretical $K_p$. Right: S/N plot for expected $K_p$ value, the grey vertical lines represent 0~km\,s$^{-1}$ and gray region represents S/N between -3 and 3.}

  \label{fig:H2O-all}
\end{figure*}

\begin{figure*}[h]
  \includegraphics[width=\textwidth]{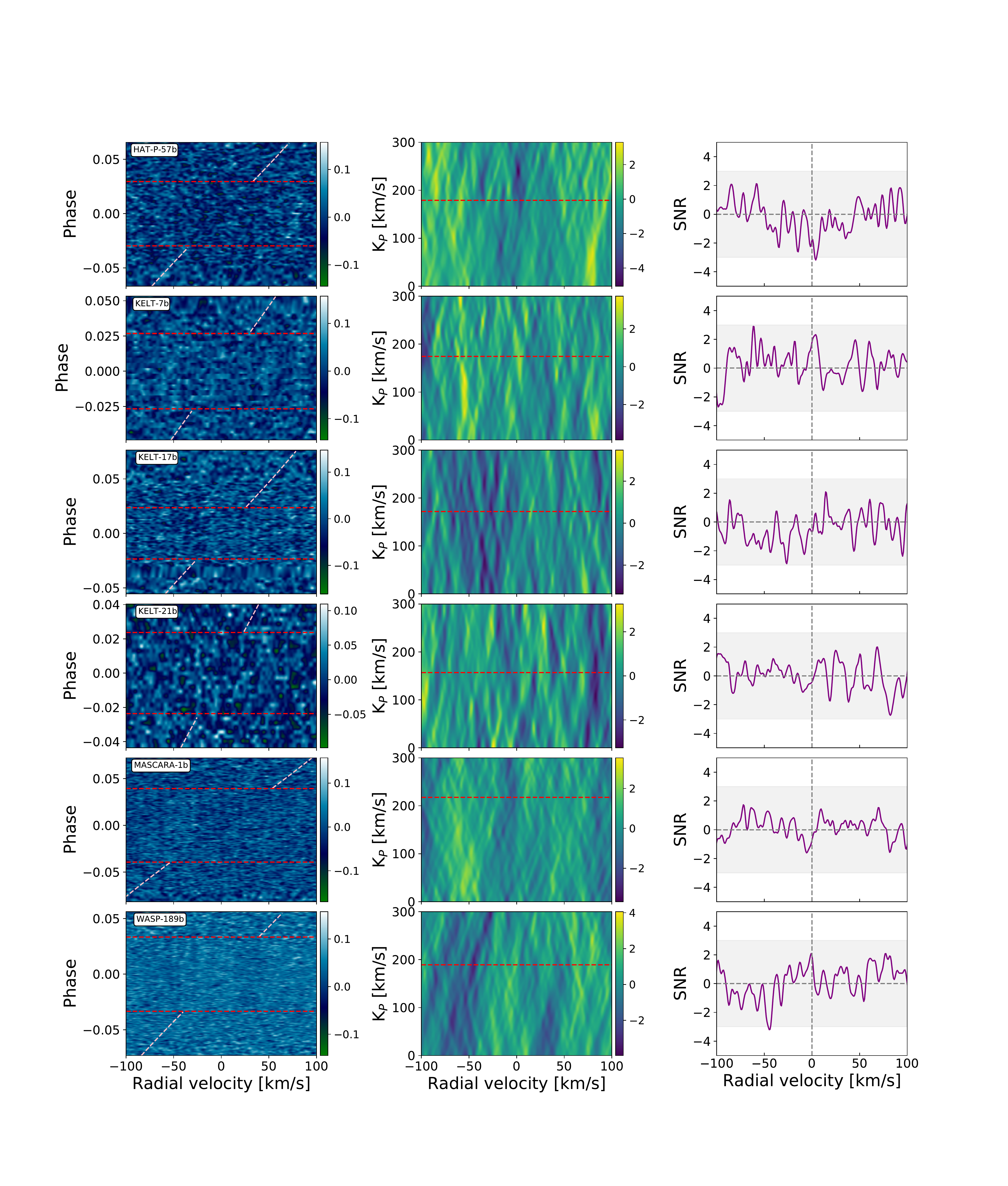}
         \caption{Cross-correlation results of \ion{K}{i} lines for the six UHJs. For each of the studied planets, we present the following plots. Left:\ Residuals map for \ion{K}{i}, where the white tilted line represents the expected velocity of the planet and red horizontal lines show the beginning and the end of the transit, region where we expect the signal from the planet; center map: $K_p$ map for $K_p$ in the range of 0 to 300~km\,s$^{-1}$ and the red line represents the theoretical $K_p$ value for the planet. We expect the signal to be coming from a planet at radial velocity of 0~km\,s$^{-1}$ and theoretical $K_p$. Right: S/N plot for expected $K_p$ value, the grey vertical lines represent 0~km\,s$^{-1}$ and gray region represents S/N between -3 and 3.}
  \label{fig:K-all}
\end{figure*}

\begin{figure*}[h]
  \includegraphics[width=\textwidth]{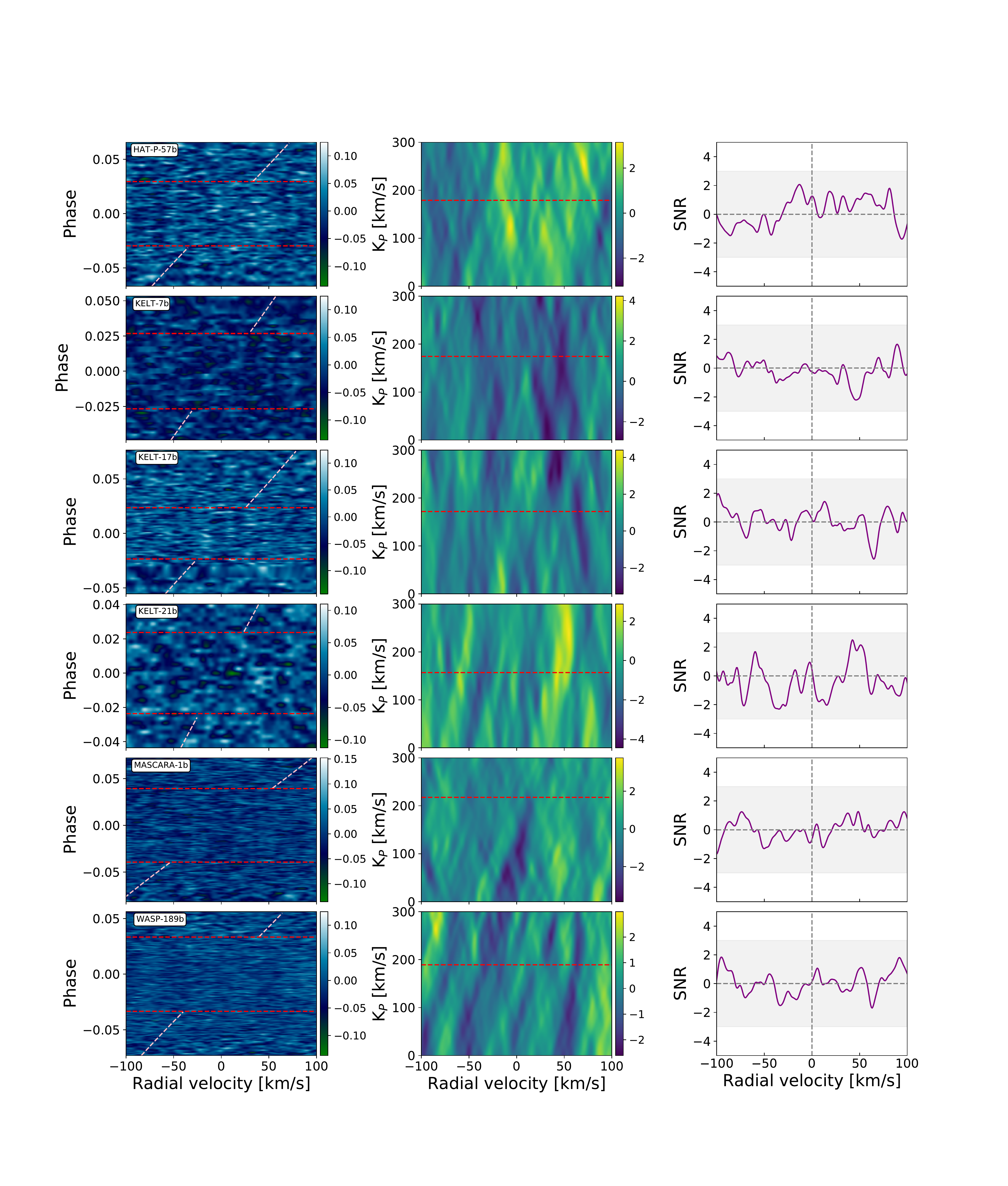}
         \caption{Cross-correlation results of \ion{Li}{i} lines for the six UHJs. For each of the studied planets, we present the following plots. Left:\ Residuals map for \ion{Li}{i}, where the white tilted line represents the expected velocity of the planet and red horizontal lines show the beginning and the end of the transit, region where we expect the signal from the planet; center map: $K_p$ map for $K_p$ in the range of 0 to 300~km\,s$^{-1}$ and the red line represents the theoretical $K_p$ value for the planet. We expect the signal to be coming from a planet at radial velocity of 0~km\,s$^{-1}$ and theoretical $K_p$. Right: S/N plot for expected $K_p$ value, the grey vertical lines represent 0~km\,s$^{-1}$ and gray region represents S/N between -3 and 3.}
  \label{fig:Li-all}
\end{figure*}

\begin{figure*}[h]
  \includegraphics[width=\textwidth]{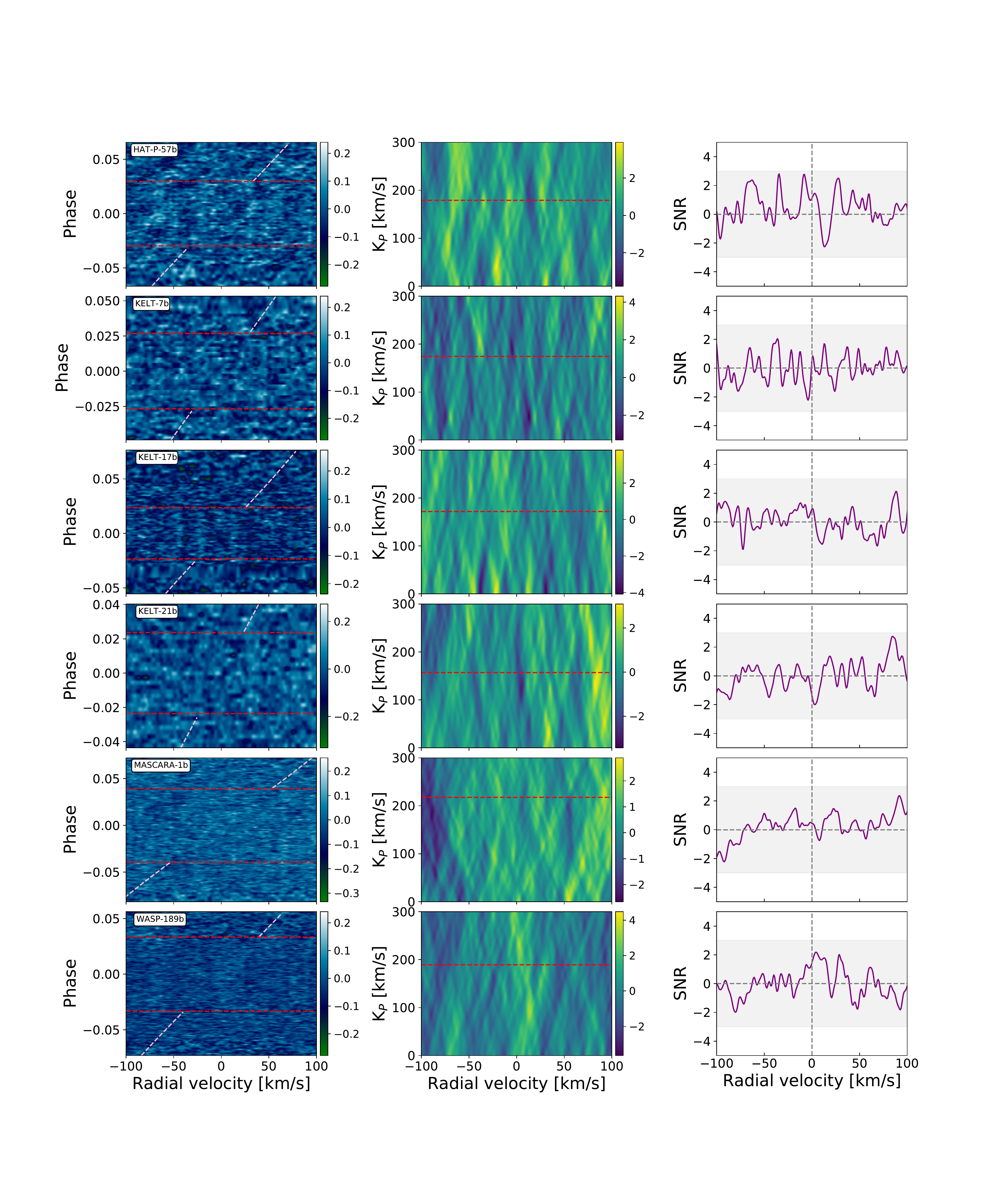}
         \caption{Cross-correlation results of \ion{Mg}{i} lines for the six UHJs. For each of the studied planets, we present the following plots. Left:\ Residuals map for \ion{Mg}{i}, where the white tilted line represents the expected velocity of the planet and red horizontal lines show the beginning and the end of the transit, region where we expect the signal from the planet; center map: $K_p$ map for $K_p$ in the range of 0 to 300~km\,s$^{-1}$ and the red line represents the theoretical $K_p$ value for the planet. We expect the signal to be coming from a planet at radial velocity of 0~km\,s$^{-1}$ and theoretical $K_p$. Right: S/N plot for expected $K_p$ value, the grey vertical lines represent 0~km\,s$^{-1}$ and gray region represents S/N between -3 and 3.}
  \label{fig:MgI-all}
\end{figure*}

\begin{figure*}[h]
  \includegraphics[width=\textwidth]{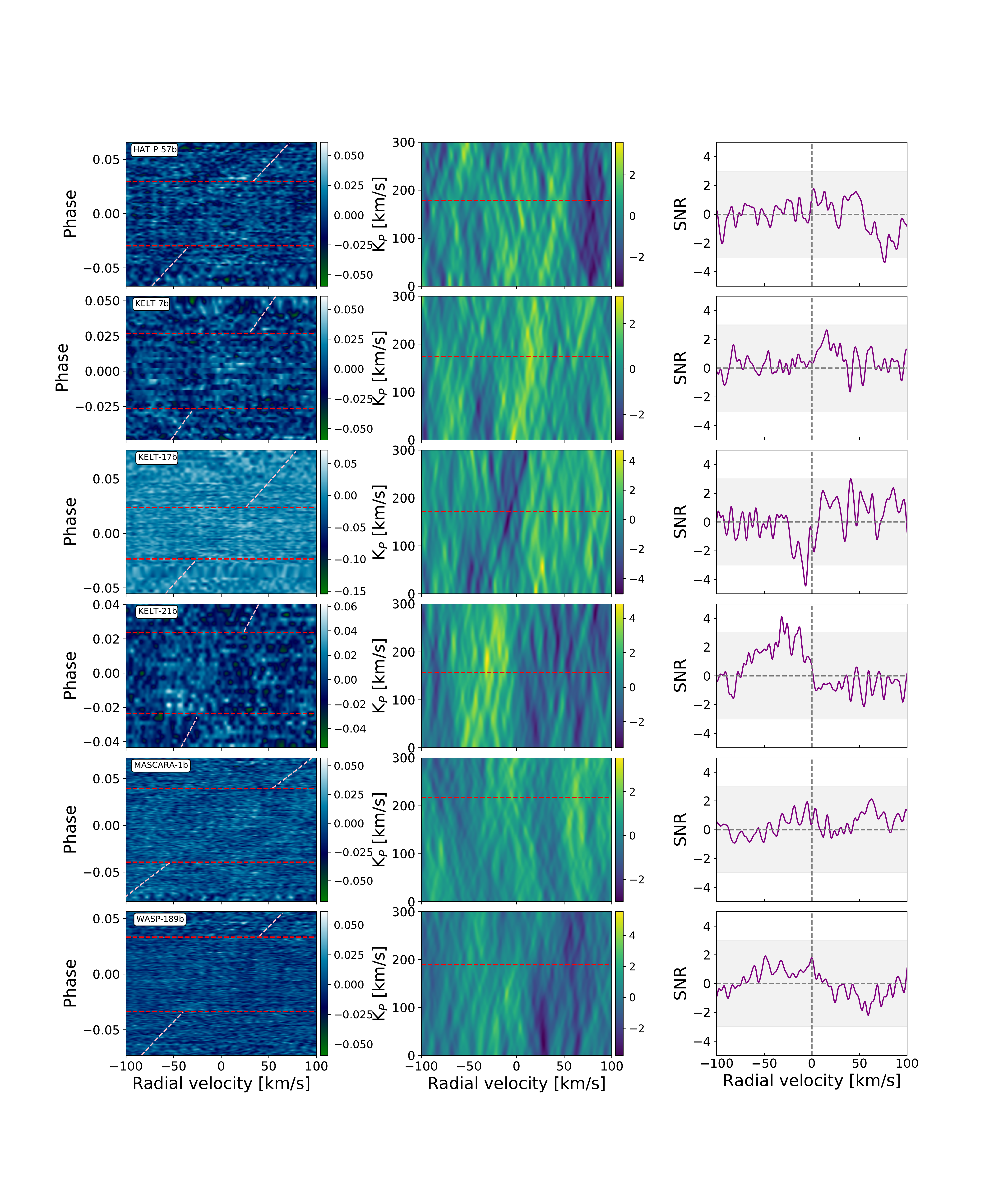}
         \caption{Cross-correlation results of \ion{Mg}{ii} lines for the six UHJs. For each of the studied planets, we present the following plots. Left:\ Residuals map for \ion{Mg}{ii}, where the white tilted line represents the expected velocity of the planet and red horizontal lines show the beginning and the end of the transit, region where we expect the signal from the planet; center map: $K_p$ map for $K_p$ in the range of 0 to 300~km\,s$^{-1}$ and the red line represents the theoretical $K_p$ value for the planet. We expect the signal to be coming from a planet at radial velocity of 0~km\,s$^{-1}$ and theoretical $K_p$. Right: S/N plot for expected $K_p$ value, the grey vertical lines represent 0~km\,s$^{-1}$ and gray region represents S/N between -3 and 3.}
  \label{fig:MgII-all}
\end{figure*}

\begin{figure*}[h]
  \includegraphics[width=\textwidth]{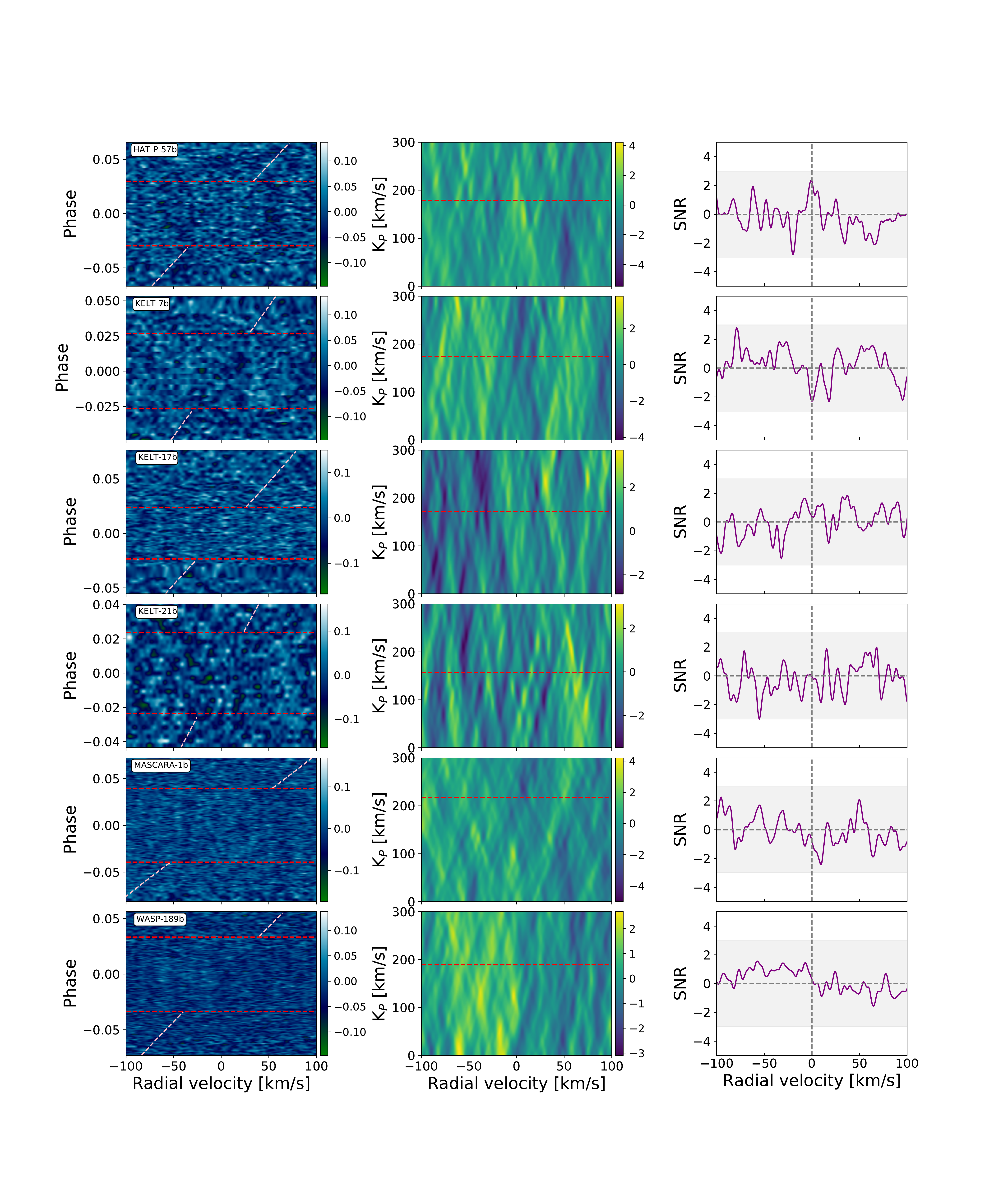}
         \caption{Cross-correlation results of \ion{Na}{i} lines for the six UHJs. For each of the studied planets, we present the following plots. Left:\ Residuals map for \ion{Na}{i}, where the white tilted line represents the expected velocity of the planet and red horizontal lines show the beginning and the end of the transit, region where we expect the signal from the planet; center map: $K_p$ map for $K_p$ in the range of 0 to 300~km\,s$^{-1}$ and the red line represents the theoretical $K_p$ value for the planet. We expect the signal to be coming from a planet at radial velocity of 0~km\,s$^{-1}$ and theoretical $K_p$. Right: S/N plot for expected $K_p$ value, the grey vertical lines represent 0~km\,s$^{-1}$ and gray region represents S/N between -3 and 3.}
  \label{fig:Na-all}
\end{figure*}

\begin{figure*}[h]
  \includegraphics[width=\textwidth]{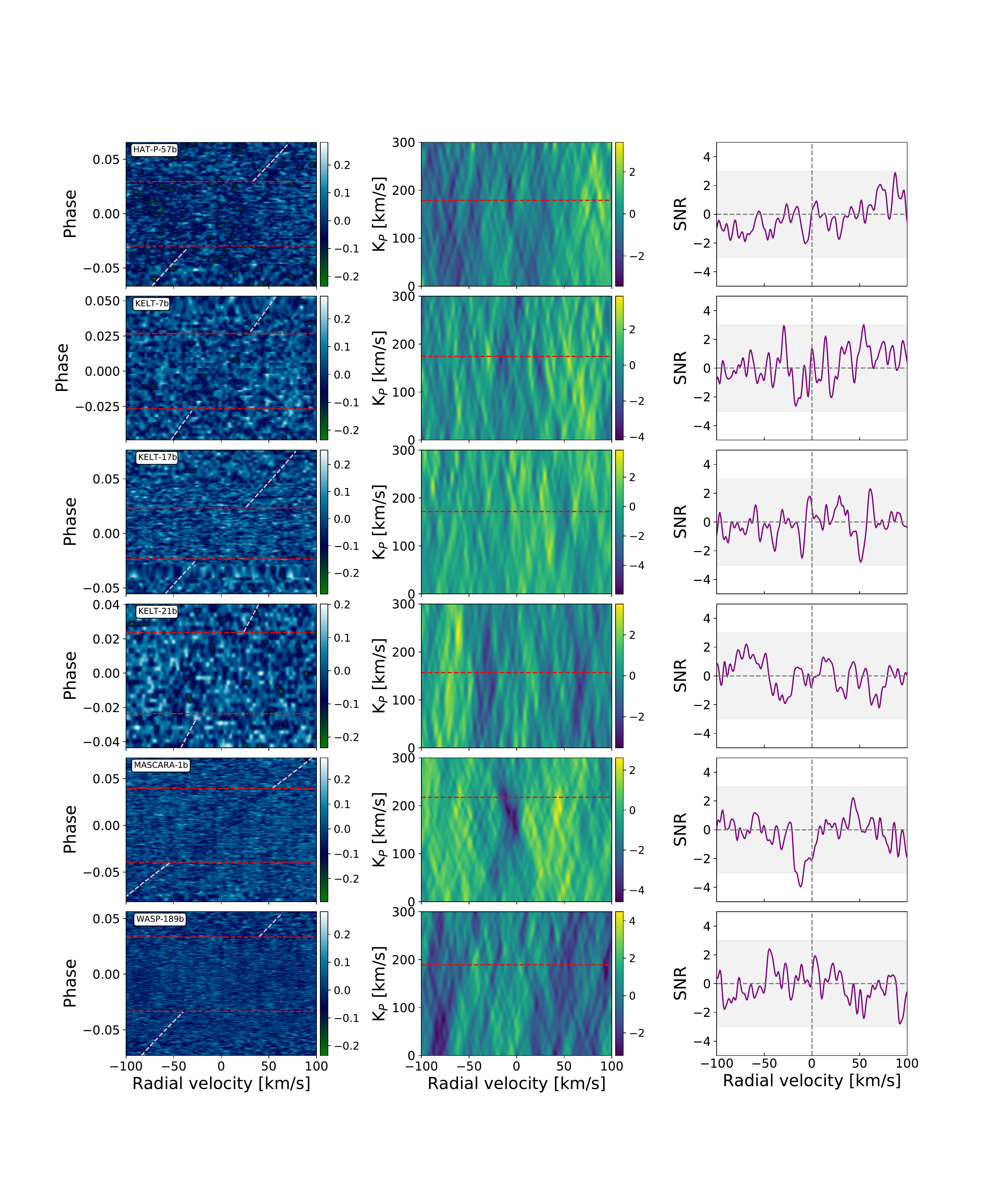}
           \caption{Cross-correlation results of \ion{Si}{i} lines for the six UHJs. For each of the studied planets, we present the following plots. Left:\ Residuals map for \ion{Si}{i}, where the white tilted line represents the expected velocity of the planet and red horizontal lines show the beginning and the end of the transit, region where we expect the signal from the planet; center map: $K_p$ map for $K_p$ in the range of 0 to 300~km\,s$^{-1}$ and the red line represents the theoretical $K_p$ value for the planet. We expect the signal to be coming from a planet at radial velocity of 0~km\,s$^{-1}$ and theoretical $K_p$. Right: S/N plot for expected $K_p$ value, the grey vertical lines represent 0~km\,s$^{-1}$ and gray region represents S/N between -3 and 3.}
  \label{fig:Si-all}
\end{figure*}

\begin{figure*}[h]
  \includegraphics[width=\textwidth]{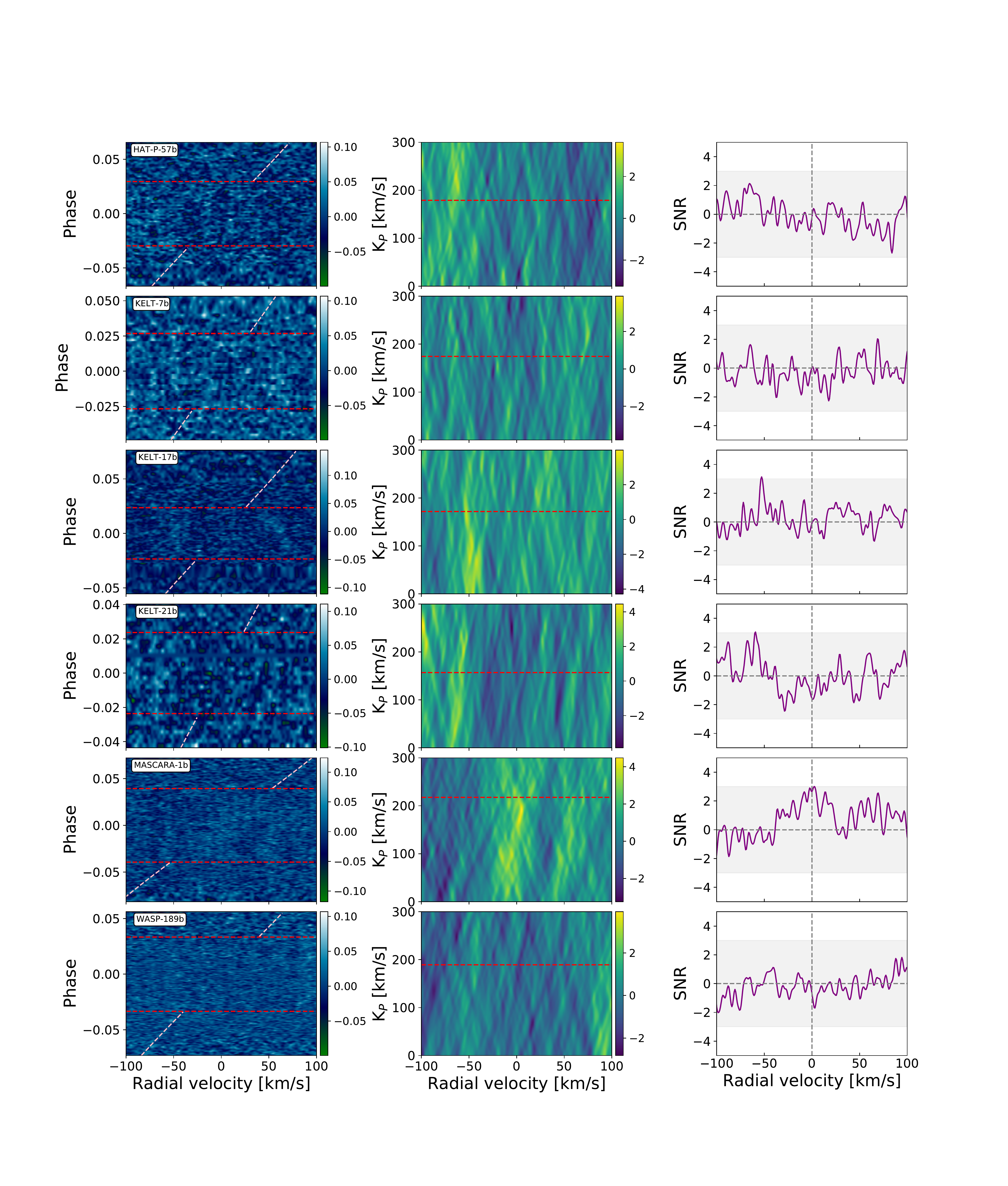}
           \caption{Cross-correlation results of \ion{Si}{ii} lines for the six UHJs. For each of the studied planets, we present the following plots. Left:\ Residuals map for \ion{Si}{ii}, where the white tilted line represents the expected velocity of the planet and red horizontal lines show the beginning and the end of the transit, region where we expect the signal from the planet; center map: $K_p$ map for $K_p$ in the range of 0 to 300~km\,s$^{-1}$ and the red line represents the theoretical $K_p$ value for the planet. We expect the signal to be coming from a planet at radial velocity of 0~km\,s$^{-1}$ and theoretical $K_p$. Right: S/N plot for expected $K_p$ value, the grey vertical lines represent 0~km\,s$^{-1}$ and gray region represents S/N between -3 and 3.}
  \label{fig:SiII-all}
\end{figure*}

\begin{figure*}[h]
  \includegraphics[width=\textwidth]{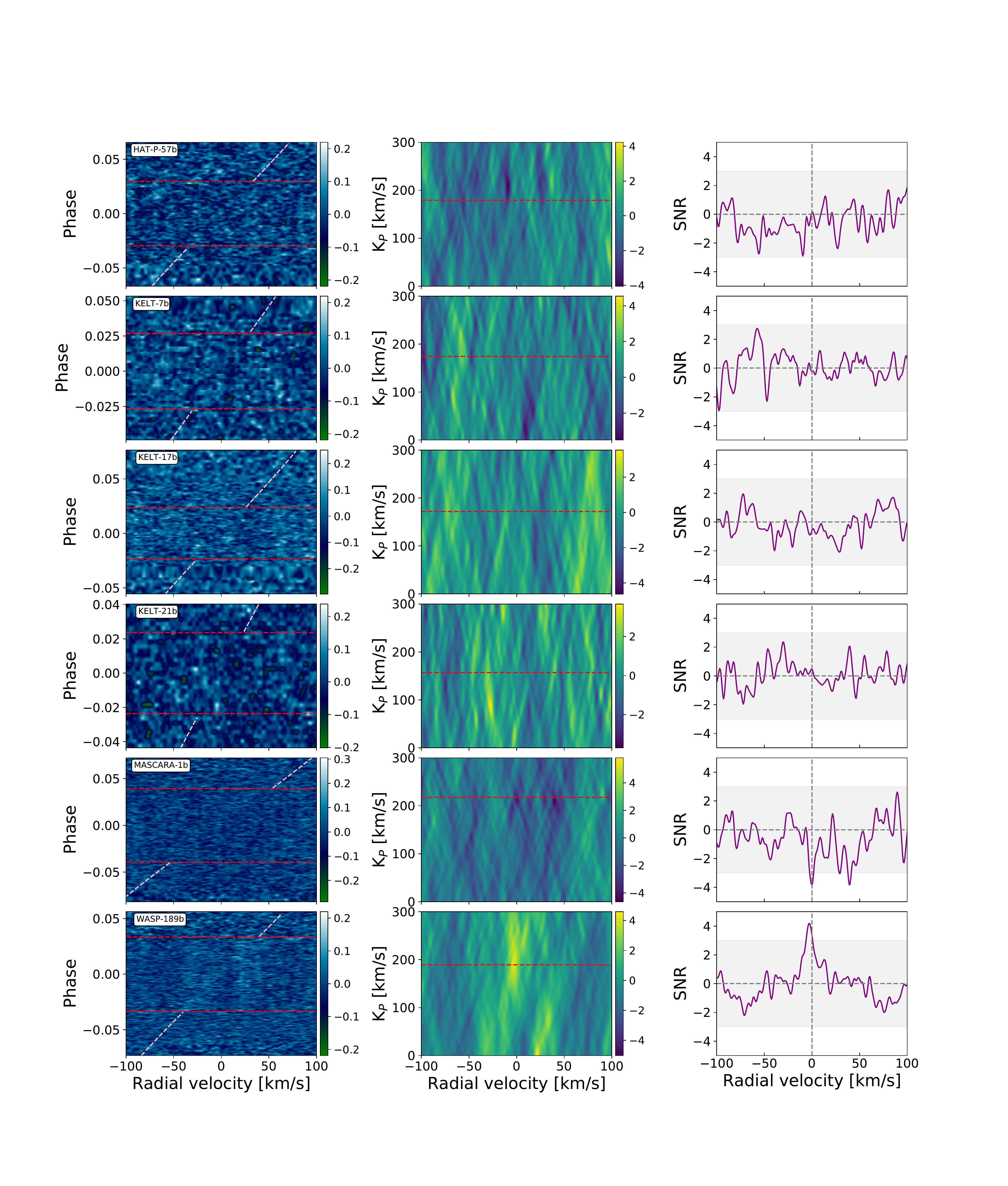}
           \caption{Cross-correlation results of \ion{Ti}{i} lines for the six UHJs. For each of the studied planets, we present the following plots. Left:\ Residuals map for \ion{Ti}{i}, where the white tilted line represents the expected velocity of the planet and red horizontal lines show the beginning and the end of the transit, region where we expect the signal from the planet; center map: $K_p$ map for $K_p$ in the range of 0 to 300~km\,s$^{-1}$ and the red line represents the theoretical $K_p$ value for the planet. We expect the signal to be coming from a planet at radial velocity of 0~km\,s$^{-1}$ and theoretical $K_p$. Right: S/N plot for expected $K_p$ value, the grey vertical lines represent 0~km\,s$^{-1}$ and gray region represents S/N between -3 and 3.}
  \label{fig:Ti-all}
\end{figure*}

\begin{figure*}[h]
  \includegraphics[width=\textwidth]{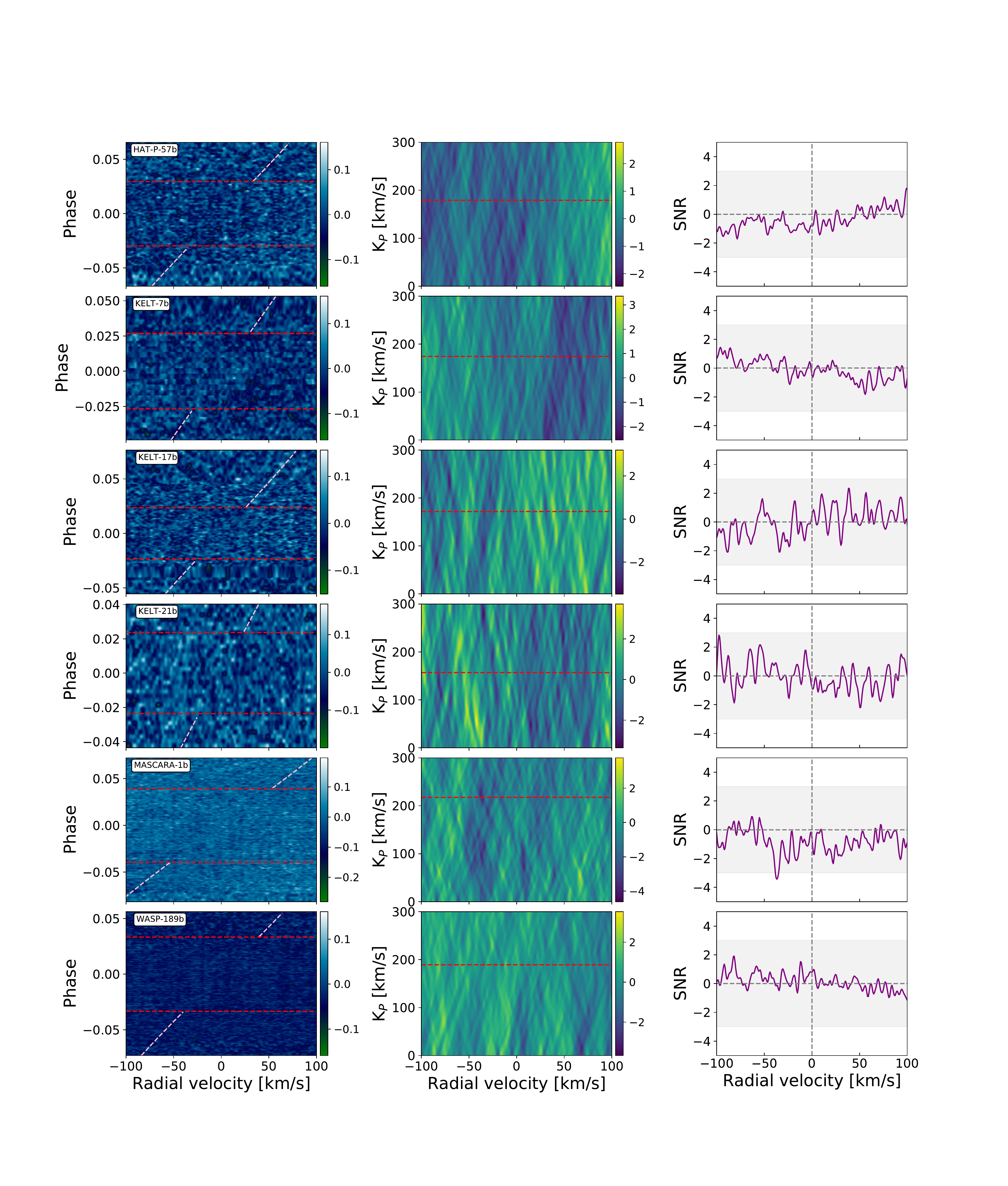}
           \caption{Cross-correlation results of TiO lines for the six UHJs. For each of the studied planets, we present the following plots. Left:\ Residuals map for TiO, where the white tilted line represents the expected velocity of the planet and red horizontal lines show the beginning and the end of the transit, region where we expect the signal from the planet; center map: $K_p$ map for $K_p$ in the range of 0 to 300~km\,s$^{-1}$ and the red line represents the theoretical $K_p$ value for the planet. We expect the signal to be coming from a planet at radial velocity of 0~km\,s$^{-1}$ and theoretical $K_p$. Right: S/N plot for expected $K_p$ value, the grey vertical lines represent 0~km\,s$^{-1}$ and gray region represents S/N between -3 and 3.}
  \label{fig:TiO-all}
\end{figure*}

\begin{figure*}[h]
  \includegraphics[width=\textwidth]{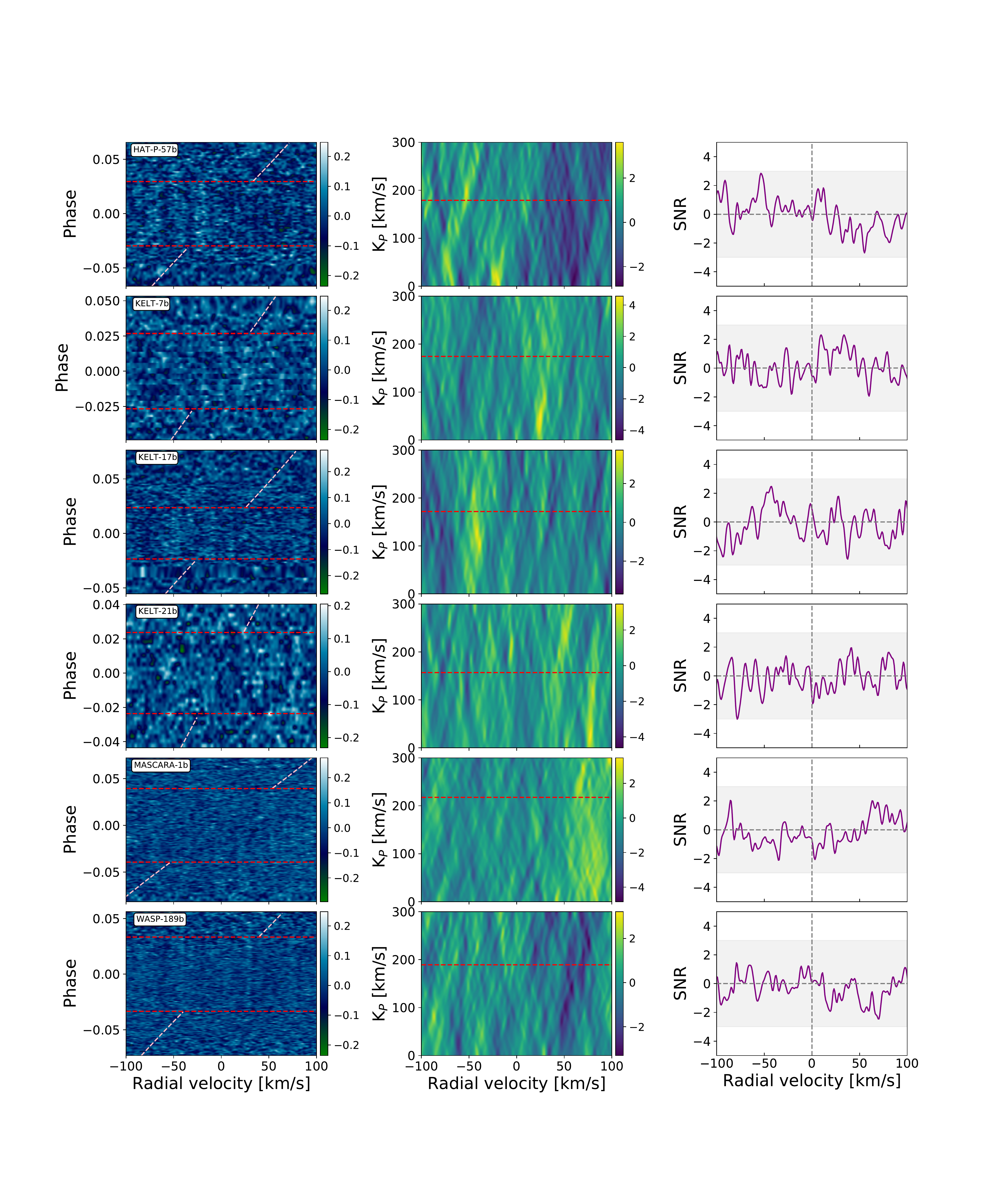}
           \caption{Cross-correlation results of \ion{V}{i} lines for the six UHJs. For each of the studied planets, we present the following plots. Left:\ Residuals map for \ion{V}{i}, where the white tilted line represents the expected velocity of the planet and red horizontal lines show the beginning and the end of the transit, region where we expect the signal from the planet; center map: $K_p$ map for $K_p$ in the range of 0 to 300~km\,s$^{-1}$ and the red line represents the theoretical $K_p$ value for the planet. We expect the signal to be coming from a planet at radial velocity of 0~km\,s$^{-1}$ and theoretical $K_p$. Right: S/N plot for expected $K_p$ value, the grey vertical lines represent 0~km\,s$^{-1}$ and gray region represents S/N between -3 and 3.}
  \label{fig:V-all}
\end{figure*}

\begin{figure*}[h]
  \includegraphics[width=\textwidth]{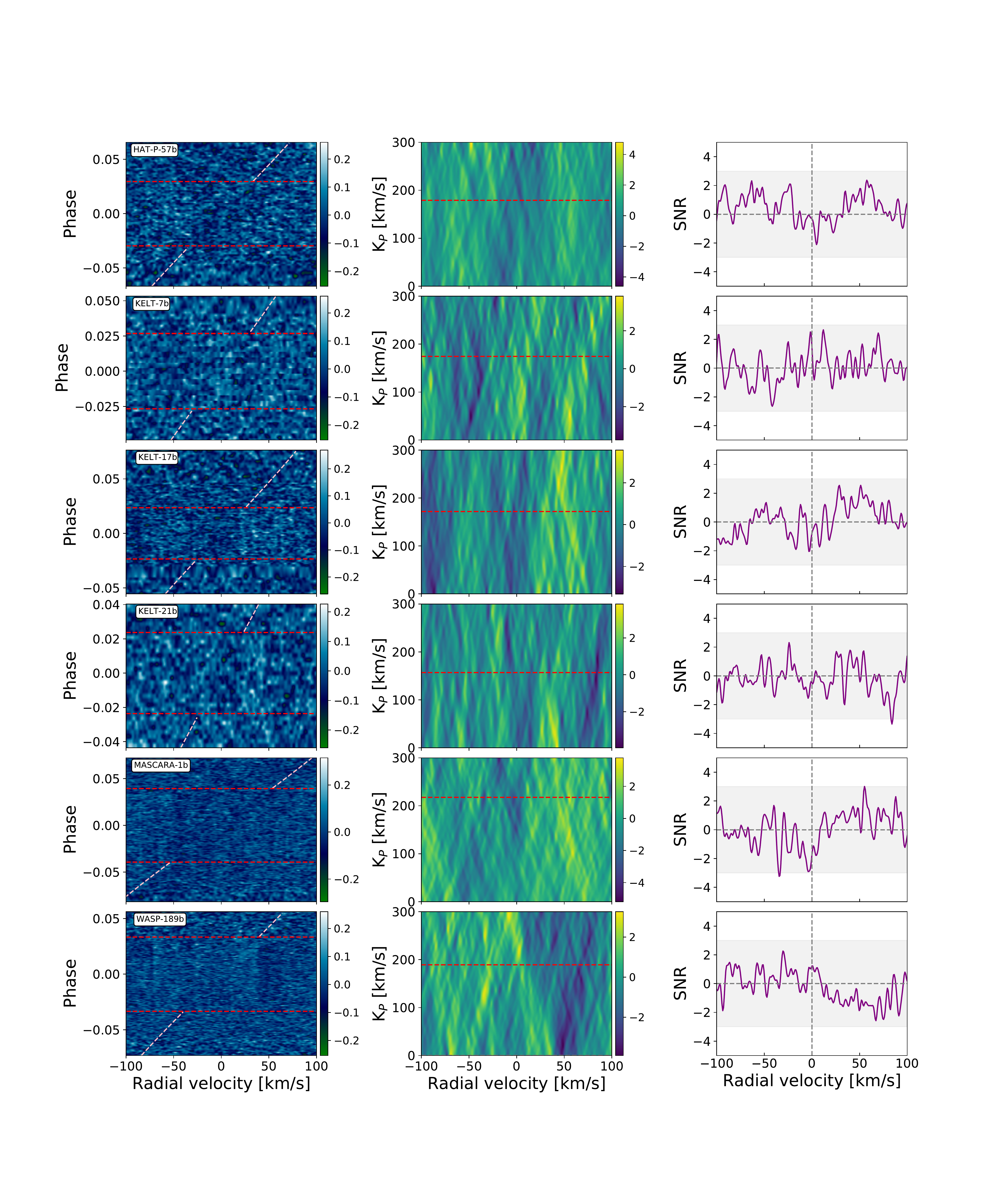}
           \caption{Cross-correlation results of \ion{V}{ii} lines for the six UHJs. For each of the studied planets, we present the following plots. Left:\ Residuals map for \ion{V}{ii}, where the white tilted line represents the expected velocity of the planet and red horizontal lines show the beginning and the end of the transit, region where we expect the signal from the planet; center map: $K_p$ map for $K_p$ in the range of 0 to 300~km\,s$^{-1}$ and the red line represents the theoretical $K_p$ value for the planet. We expect the signal to be coming from a planet at radial velocity of 0~km\,s$^{-1}$ and theoretical $K_p$. Right: S/N plot for expected $K_p$ value, the grey vertical lines represent 0~km\,s$^{-1}$ and gray region represents S/N between -3 and 3.}
  \label{fig:VII-all}
\end{figure*}

\begin{figure*}[h]
  \includegraphics[width=\textwidth]{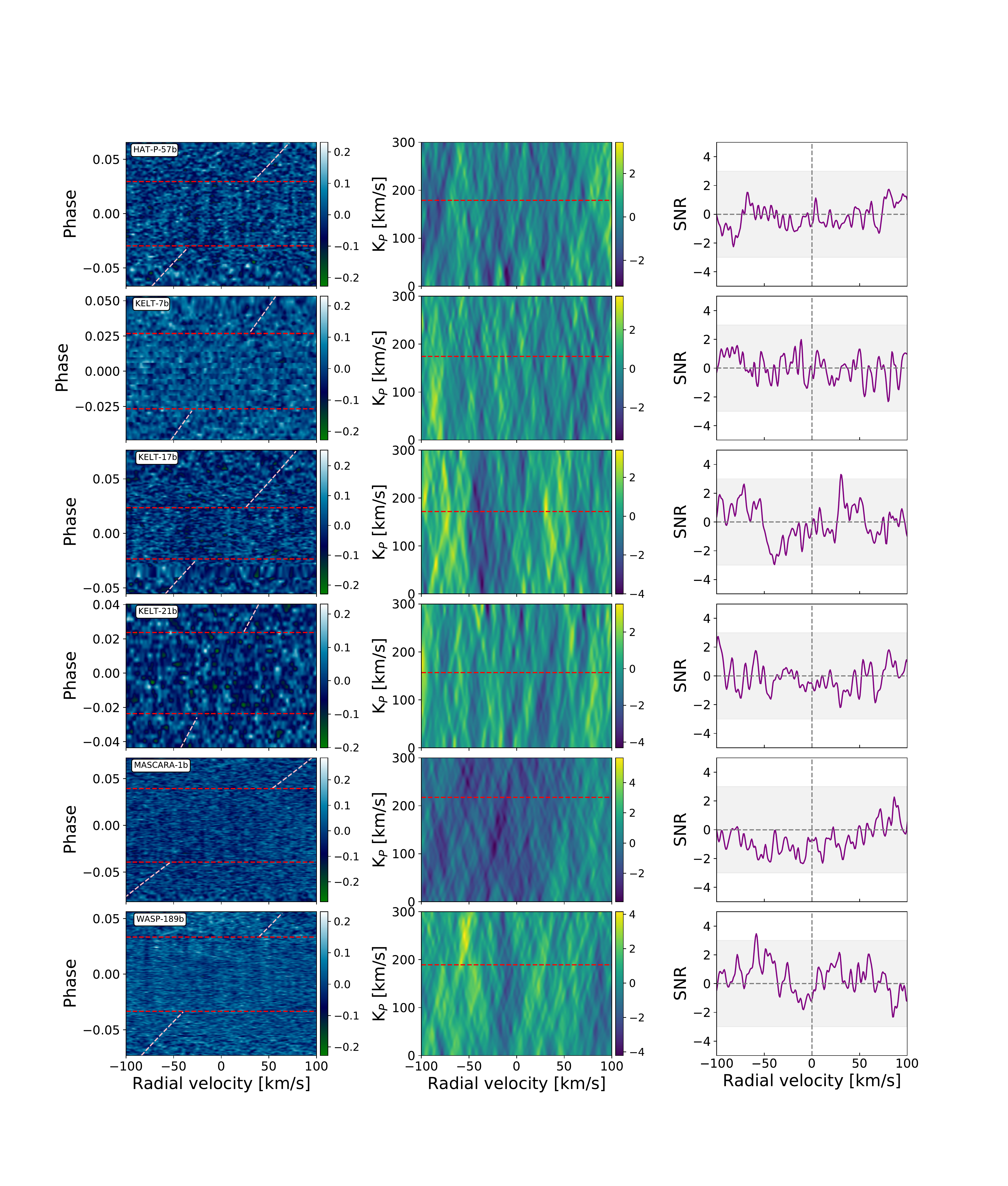}
           \caption{Cross-correlation results of VO lines for the six UHJs. For each of the studied planets, we present the following plots. Left:\ Residuals map for VO, where the white tilted line represents the expected velocity of the planet and red horizontal lines show the beginning and the end of the transit, region where we expect the signal from the planet; center map: $K_p$ map for $K_p$ in the range of 0 to 300~km\,s$^{-1}$ and the red line represents the theoretical $K_p$ value for the planet. We expect the signal to be coming from a planet at radial velocity of 0~km\,s$^{-1}$ and theoretical $K_p$. Right: S/N plot for expected $K_p$ value, the grey vertical lines represent 0~km\,s$^{-1}$ and gray region represents S/N between -3 and 3.}
  \label{fig:VO-all}
\end{figure*}

\begin{figure*}[h]
  \includegraphics[width=\textwidth]{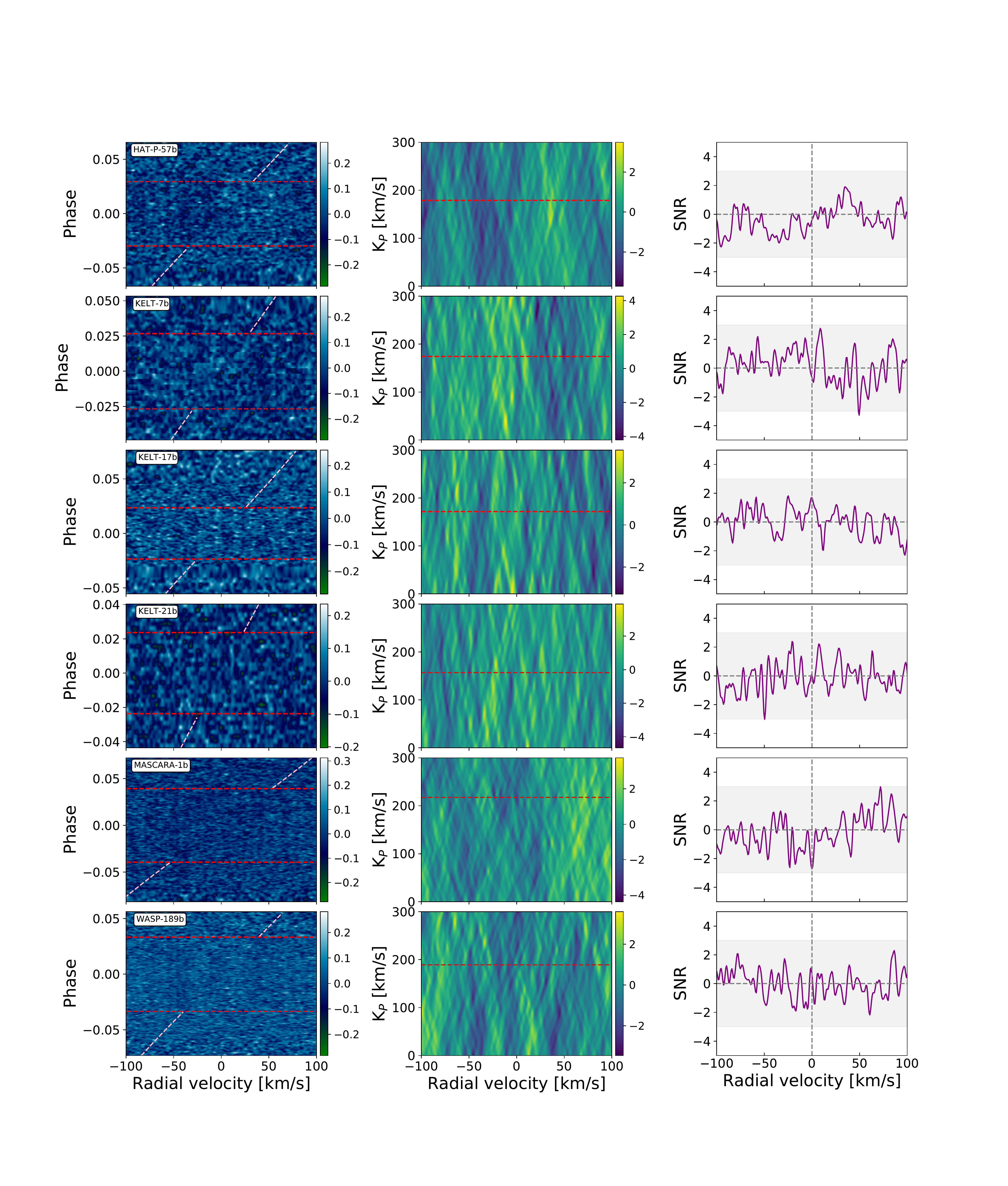}
           \caption{Cross-correlation results of \ion{Y}{i} lines for the six UHJs. For each of the studied planets, we present the following plots. Left:\ Residuals map for \ion{Y}{i}, where the white tilted line represents the expected velocity of the planet and red horizontal lines show the beginning and the end of the transit, region where we expect the signal from the planet; center map: $K_p$ map for $K_p$ in the range of 0 to 300~km\,s$^{-1}$ and the red line represents the theoretical $K_p$ value for the planet. We expect the signal to be coming from a planet at radial velocity of 0~km\,s$^{-1}$ and theoretical $K_p$. Right: S/N plot for expected $K_p$ value, the grey vertical lines represent 0~km\,s$^{-1}$ and gray region represents S/N between -3 and 3.}
  \label{fig:Y-all}
\end{figure*}

\end{appendix}

\end{document}